\documentstyle[12pt]{article}
\newcommand{\secn}[1]{Section~\ref{#1}}
\newcommand{\bra}[1]{\langle{#1}|}
\newcommand{\ket}[1]{|{#1}\rangle}

\newcommand{\eq}[1]{Eq.~(\ref{#1})}

\def\beq{\begin{equation}}
\def\eeq{\end{equation}}
\def\beqa{\begin{eqnarray}}
\def\eeqa{\end{eqnarray}}
\newcommand{\sect}[1]{\setcounter{equation}{0}\section{#1}}
\renewcommand{\theequation}{\thesection.\arabic{equation}}
\newcommand{\EQ}{\begin{equation}}
\newcommand{\EN}{\end{equation}}
\newcommand{\bea}{\begin{eqnarray}}
\newcommand{\ena}{\end{eqnarray}}

\renewcommand{\a}{\alpha}
\renewcommand{\b}{\beta}

\newcommand{\e}{\epsilon}

\newcommand{\NP}[1]{Nucl.\ Phys.\ {\bf #1}}
\newcommand{\PL}[1]{Phys.\ Lett.\ {\bf #1}}

\newcommand{\PRL}[1]{Phys.\ Rev.\ Lett.\ {\bf #1}}

\renewcommand{\thefootnote}{\fnsymbol{footnote}}
\def\one{{\hbox{ 1\kern-.8mm l}}}
\def\gh{{\rm gh}}
\def\sgh{{\rm sgh}}
\def\NS{{\rm NS}}
\def\R{{\rm R}}
\def\ii{{\rm i}}
\def\bz{{\bar z}}
\def\comm#1#2{\left[ #1, #2\right]}
\def\acomm#1#2{\left\{ #1, #2\right\}}
\def\tr{{\rm tr\,}}
\newlength{\bredde}
\def\slash#1{\settowidth{\bredde}{$#1$}\ifmmode\,\raisebox{.15ex}{/}
\hspace*{-\bredde} #1\else$\,\raisebox{.15ex}{/}\hspace*{-\bredde} #1$\fi}
\begin{document}
\begin{titlepage}
\rightline{DFTT 6/98}
\rightline{NORDITA 98/10-HE}
\rightline{KUL-TF-98/10}
%\rightline{hep-th/9802088}
\rightline{\hfill February 1998}
\vskip 1.2cm
%\vskip 0.8cm
\centerline{\Large \bf Microscopic string analysis of the } 
\centerline{\Large \bf D0-D8 brane system and dual 
R-R states\footnote{Work partially supported by the European Commission
TMR programme ERBFMRX-CT96-0045 in which R.R. is associated to
Torino University, and by MURST.}}
%\vskip 1.2cm
\vskip 0.8cm
\centerline{\bf M. Bill\'o$^a$\footnote{e-mail:
billo@tfdec1.fys.kuleuven.ac.be}, P. Di Vecchia$^b$, M. Frau$^{c,f}$,
A. Lerda$^{d,c,f}$,}
\vskip .2cm
\centerline{\bf I. Pesando$^{c,f}$, R. Russo$^{e,f}$ and S. Sciuto$^{c,f}$}
\vskip .5cm
\centerline{\sl $^a$ Instituut voor theoretische fysica,}
\centerline{\sl Katholieke Universiteit Leuven, B-3001 Leuven, Belgium}
\vskip .2cm
\centerline{\sl $^b$ NORDITA, Blegdamsvej 17, DK-2100 Copenhagen \O, Denmark}
\vskip .2cm
\centerline{\sl $^c$ Dipartimento di Fisica Teorica, Universit\`a di
Torino}
\vskip .2cm
\centerline{\sl $^d$ Dipartimento di Scienze e Tecnologie Avanzate}
\centerline{\sl Universit\`a di Torino, sede di Alessandria}
\vskip .2cm
\centerline{\sl $^e$ Dipartimento di Fisica, Politecnico di Torino}
\vskip .2cm
\centerline{\sl $^f$ I.N.F.N., Sezione di Torino, Via P. Giuria 1, I-10125 
Torino, Italy}
\vskip 0.8cm
\begin{abstract}
%\centerline{\bf Abstract}
%\small
Using the boundary state formalism, we perform a microscopic string
analysis of the interaction between two D-branes and provide
a local interpretation for the R-R force in the D0-D8 brane system.
To do so, we construct  
BRST invariant vertex operators for the massless R-R states in the
asymmetric picture that are proportional to potentials  
rather than field strengths.
The Hilbert space of such R-R states contains combinations of
two vectors that decouple
from all physical amplitudes, even in the presence of boundaries. 
Identifying these vectors, we remove the null states and 
recover duality relations among  R-R potentials. 
If we specify to the D0-D8 brane system, this mechanism implies
that the R-R $1$-form state has a non-zero overlap 
with both the D0-brane and the D8-brane, thus
explaining from a local point of view the non-vanishing R-R contribution 
in the interaction for the
D0-D8 brane system and those related to it by duality.
%\normalsize
\end{abstract}
\end{titlepage}
\newpage
\renewcommand{\thefootnote}{\arabic{footnote}}
\setcounter{footnote}{0}
\setcounter{page}{1}
\sect{Introduction}
\label{intro}
%\vskip 0.5cm
After the discovery \cite{POLCH} that D-branes are characterized by the 
fact that open strings with Dirichlet boundary conditions
can end on them, it has been possible to study the interaction between
two D-branes by computing a one-loop open string diagram with the 
open string stretching between the branes \cite{POLCH,LPOLCH}.
An alternative way of describing D-branes is provided by the
boundary state \cite{NBOUND,BILLO,FRAU,OTHERS,GABE}. 
This is a BRST invariant state which
can be interpreted as a source for a closed string emitted by a
D-brane, and as such it is directly related to the classical
brane solutions of the low-energy string effective action
\cite{cpb}. From this standpoint, the interaction between two D-branes
is viewed as an exchange of closed string states, and thus it 
is computed with a tree-level diagram in which two
boundary states are connected to each other by means of
a closed string propagator.
These two apriori completely independent approaches are actually
equivalent and give exactly the same results,
as a consequence of the modular properties of the  
string diagrams. From the explicit expression of the amplitude, one
can easily see that when the two D-branes are near to each other,
only the massless open string states are responsible for the
interaction, whereas in the case of two distant D-branes only the massless
closed string states give a non vanishing contribution.

If the two D-branes form a configuration which preserves enough 
space-time supersymmetry, then
they do not interact. For example, this happens with two parallel 
D$p$-branes, which break half of the supersymmetries. In this case,
the vanishing of the force at the string level is a consequence of the 
``abstruse identity'' and, from a field theory point of view, it can
be understood~\cite{LPOLCH} as due to a cancellation between the
attractive contribution provided by the 
graviton and dilaton exchanges in the NS-NS sector, and the repulsive 
contribution provided by the exchange of a $(p+1)$-form potential in 
the R-R sector. A zero force between two D-branes is also found in all
other configurations in which the number $\nu$ of mixed 
Dirichlet-Neumann conditions for the open string stretching between 
the two D-branes is equal to $4$ or $8$~\cite{POLCH,LPOLCH,LIF}. 
The case $\nu=4$ has an explanation similar to the one we have just 
mentioned for parallel branes ($\nu=0$). At the string level 
both the NS-NS and the R-R sectors are separately vanishing, and 
this is consistent with the fact that at large distance 
the graviton and the dilaton exactly compensate each other, while 
there is no massless R-R field which can be exchanged between  
the two D-branes since these couple to different R-R states.

On the contrary, the case $\nu=8$ is more difficult to understand. 
At the string level, the vanishing of the force is again due to the
``abstruse identity'' which provides a cancellation between the 
non-zero contributions of the NS-NS and R-R sectors, but its 
microscopic interpretation is in this case problematic. In fact, 
the graviton and dilaton exchanges yield a repulsive force which
should be compensated by an attractive one coming
from the R-R potentials. However, from the field theory point of view,
the branes of a $\nu=8$ system seem to couple to different R-R 
states, and thus it appears impossible that they can exchange
a R-R field. On the other hand, it is undoubted that, when $\nu=8$,
there is a non-vanishing contribution from the massless R-R sector,
which has also been used to explain the
anomalous string creation when, for example, a D$0$-brane passes 
adiabatically through a D$8$-brane~\cite{GABE,DANI,GBD}. In fact,
it turns out that after the two branes have exchanged their relative 
positions, the R-R force increases by a quantity proportional to the
string tension $T=1/(2 \pi \alpha ')$.
This effect can then be interpreted as due to the creation
of a fundamental string, and is related by a sequence of dualities
to the creation of branes of different dimensions in other systems
~\cite{HW}.

The puzzle of the field theory interpretation of the string 
results in the $\nu=8$ system has been addressed and discussed in
Ref.~\cite{DANI}, where the 
unexpected R-R contribution is explained by treating the R-R 
1-form emitted by a D0-brane as a background field for a D8-brane. 
Thus, the total R-R force can be directly read from the 
effective lagrangian as a non-local effect.
An alternative explanation can be obtained by looking at  
the structure of the action 
of massive low-energy IIA supergravity~\cite{GABE,STROPO}.
However, all these arguments are not on the same footing as the 
ones already discussed for the cases $\nu=0, 4$, 
where  the contributions of the various string sectors, that conspire to
give the vanishing total result, are understood in terms of the 
exchanges of a dilaton, a graviton and a R-R state. 

The main result of this paper is to show that also for the
$\nu=8$ systems it is possible to explain the R-R interaction 
from a microscopic and local point of view 
just like one usually does in all other cases.
The crucial point is that the R-R charges
of the two D-branes of a $\nu=8$ system 
are essentially identified by a duality relation, and produce the same
R-R potential. Therefore, the R-R interaction is nothing but 
the usual Coulomb-like force between D-branes. 

More precisely in this paper we use the BRST invariant
expression of the boundary state 
to compute, in the covariant formalism, the interaction between two 
D-branes as a closed string tree-level
diagram. The contribution of the zero modes of the R-R sector turns out
to be ill defined and hence, following Ref.~\cite{YOST}, we introduce a 
regulator to obtain a meaningful result. In this way, we 
clearly see that the divergent superghost part and the vanishing
matter contribution, when put together in the full amplitude,
combine to leave a finite non-zero result if
$\nu =8$, while they give a vanishing result in all other cases.
The final expression of the amplitude between two D-branes obtained in 
this way agrees with the one computed in Refs.~\cite{GABE,LIF} without
introducing the superghosts.

To give a microscopic interpretation of the force between two
D-branes, it is necessary to know which closed string states
can couple to the boundary state. While in the NS-NS sector
there are no particular problems, it was already observed a few
years ago in  Ref.~\cite{sagnotti}, that a boundary state has a non
zero overlap only with R-R states in an asymmetric picture where
the correspondent vertex operators contain the R-R potentials.
On the contrary, the R-R states that are usually considered in 
perturbative string theory are in a symmetric picture 
and are proportional to the field strengths of the R-R potentials.
To overcome this problem, we have explicitly
constructed  BRST invariant vertex operators 
for massless R-R states in 
the asymmetric picture, and found that they are proportional to the R-R
potentials, as suggested few years ago in  
Ref.~\cite{sagnotti}, rather than to the corresponding field strengths.
Furthermore, since the massless states exchanged between two
D-branes cannot be on shell, it
is necessary to study their off-shell properties.
For the NS-NS sector it is known that the propagating states emitted by 
a D-brane are in the cohomology of a suitably restricted charge $Q'$
\cite{RamThor}. We have found that for the R-R sector in the 
asymmetric picture there is a unique class of vertices $W$ 
that are invariant under the restricted charge $Q'$ also when extended
off-shell. These new vertex operators contain 
an infinite number of terms, do not have left 
and right superghost number separately defined, and create states 
$\ket{W}$ that have exactly the same structure of the zero-mode part of 
the boundary states of the R-R sector.
Despite the presence of infinite terms, these new states
have a well-defined norm, provided that
the scalar product is defined using the same 
regularization prescription introduced for boundary states. 
With this definition, a state $\ket{W}$ has a 
non vanishing scalar product not only with itself, but also 
with states $\ket{W'}$ carrying forms of different degree.

A careful analysis shows that there are two dimensional subspaces of
the R-R Hilbert space with a degenerate metric. Thus, in each one
of these subspaces, there exists a combination of two vectors 
which is null state, {\em i.e.} a state decoupling
from all amplitudes, even with boundaries.
This fact shows that in the Hilbert space of the asymmetric
R-R sector there are pairs of dual vectors which describe the same
state. On shell this is the
usual Hodge duality between R-R potentials, whereas
if we specify to the D0-D8 brane system, we see 
that, because of these identifications, 
the R-R $1$-form state has a non-zero
overlap with both the D$0$-brane and the D$8$-brane, thus solving the 
above-mentioned problem of the non vanishing R-R contribution to the
interaction.

The paper is organized as follows. In \secn{boundary}, we give the full 
expression of the BRST invariant boundary state. \secn{interaction} is 
devoted to the calculation of the static interaction between two 
D-branes. In \secn{RRstate} we construct the BRST invariant vertex 
operators in the asymmetric picture describing massless R-R potentials. 
In \secn{factor} we derive the duality relations between 
different R-R potentials and use them to show that for example a
D0 and a D8 brane have opposite R-R charges. The factorization of the
brane amplitude and its microscopic 
interpretation then follow immediately. Finally, in Appendix A we give 
some details on the
construction of the BRST invariant asymmetric vertex operators
for the R-R massless states with indefinite superghost number, and in 
Appendix B we show their BRST equivalence with vertex operators of zero
left and right superghost number. 
%\vskip 1cm
\sect{Boundary state for a D$p$-brane}
\label{boundary}
%\vskip 0.5cm
As explained in Ref. \cite{cpb} \footnote{For an earlier discussion of the
boundary state for the pure Neumann case see Refs.~\cite{YOST,callan}.},
the boundary state $\ket{B}$ is a BRST invariant state of the
closed string that inserts a boundary on the world-sheet and
enforces the boundary conditions appropriate for a D-brane.
For both the NS-NS and R-R sectors of the fermionic
string, $\ket{B}$ can be written as the product of
a matter part and a ghost part
\begin{equation}
\label{bs0}
\ket{B} = \ket{B_{\rm mat}} \ket{B_{\rm g}}~~,
\end{equation}
where
\begin{equation}
\ket{B_{\rm mat}} = \ket{B_X} \ket{B_{\psi}}~~~,~~~
\ket{B_{\rm g}} = \ket{B_{\rm gh}} \ket{B_{\rm sgh}}~~.
\label{bs00}
\end{equation}
The matter part $\ket{B_{\rm mat}}$
is defined by the overlap conditions
that fix the identification at the
boundary between the left and right movers of the
matter fields $X^\mu$ and $\psi^\mu$, namely
\begin{eqnarray}
\partial_\tau X^\alpha|_{\tau=0} \ket{B_X } = 0
~~~&,&~~~ (X^i - y^i)|_{\tau=0}\ket{B_X} = 0
~~, \label{bs2} \\
(\psi^\alpha-\ii\eta \tilde\psi^\alpha)|_{\tau=0}\ket{B_{\psi},\eta}
= 0
~~~&,&~~~
(\psi^i+\ii\eta \tilde\psi^i)|_{\tau=0}\ket{B_{\psi},\eta} = 0~~.
\label{bs2bis}
\end{eqnarray}
where $\alpha$ labels the $(p+1)$ Neumann (or longitudinal)
directions,
$i$ labels the $(9-p)$ Dirichlet (or transverse)
directions of a D$p$-brane located at $y$.
Notice that for $\psi^\mu$ there are two consistent
identifications (corresponding to $\eta=\pm1$), but, as we shall see,
the GSO projection will allow only a
superposition of the two.
\par
By introducing  the matrix
\begin{equation}
S_{\mu \nu} = ( \eta_{\alpha \beta} , - \delta_{ij} )   ~~,
\label{smunu}
\end{equation}
and expanding the fields in modes,
the overlap relations (\ref{bs2}) and (\ref{bs2bis}) become respectively
\bea
&& \Big( \a_{n} + S \cdot {\tilde{\a}}_{-n}\Big)\ket{B_X}
=0 ~~~ (n \neq 0)~~,\nonumber \\
&& \hat p^{\alpha} \ket{B_X} = ( {\hat q}^i - y^i ) \ket{B_X} = 0~~,
\label{xcond}
\ena
and
\begin{equation}
\left( \psi_{m} - \ii \eta S \cdot {\tilde{\psi}}_{-m}\right)
\ket{B_{\psi}, \eta }
=0
\label{psicond}
\end{equation}
where the index $m$ is integer in the R sector and half-integer
in the NS sector.
\par
It is not difficult to check that the
identifications (\ref{xcond}) and (\ref{psicond})
imply that $\ket{B_{\rm mat}, \eta}$ is annihilated by the following
linear combinations of left and right generators of
the super Virasoro algebra
\begin{equation}
\left( { L}_n^{\rm mat} - {\tilde{{ L}}}_{-n}^{\rm mat} \right)
\ket{B_{\rm mat}, \eta} =0 ~~~,~~~
\left( { G}_m^{\rm mat} + \ii \eta  {\tilde{G}}_{-m}^{\rm mat} \right)
\ket{B_{\rm mat}, \eta} =0~~.
\label{viraco}
\end{equation}
Since the boundary state $\ket{B, \eta}$ must be BRST invariant,
that is\footnote{The boundary state must be
in the cohomology of the total BRST charge $Q+\tilde Q$; contrarily
to the usual perturbative states of the closed string,
it cannot be chosen to be annihilated
separately by $Q$ and $\tilde Q$ due to the presence of
a boundary on the world-sheet.}
\begin{equation}
\label{bs1}
\left({ Q} + { \tilde Q}\right)\ket{B,\eta} = 0~~,
\end{equation}
the relations (\ref{viraco}) must be supplemented by the
analogous ones in the ghost sector, namely
\begin{equation}
\left( {L}_{n}^{\rm g} - {\tilde{L}}_{-n}^{\rm g} \right)
\ket{B_{\rm g}, \eta} =0
~~~,~~~
\left({G}_{m}^{\rm g} + \ii \eta  {\tilde{{ G}}}_{-m}^{\rm g}
\right) \ket{B_{\rm g}, \eta} =0  ~~.
\label{lngh}
\end{equation}
They imply that
\begin{eqnarray}
\label{bsghost}
\left(c_{n} + {\tilde{c}}_{-n}\right) \ket{B_{\rm gh} } = 0~~~~, & &
\left(b_{n} - {\tilde{b}}_{-n}\right) \ket{B_{\rm gh}} = 0~~,
\nonumber\\
\left(\gamma_{m} +\ii\eta {\tilde{\gamma}}_{-m}\right)
\ket{B_{\rm sgh} ,\eta} = 0~~~~,
& &
\left(\beta_{m} +\ii \eta {\tilde{\beta}}_{-m}\right)
\ket{B_{\rm sgh} ,\eta} =
0~~.
\end{eqnarray}
\par
Using the conventions and
normalizations of Ref. \cite{cpb}, we can write
the solution to the overlap equations
(\ref{xcond}), (\ref{psicond}) and (\ref{bsghost}) as follows
\begin{equation}
\label{bs3}
\ket{B,\eta}_{\rm R,NS} = {T_p\over 2}
\ket{B_X}\, \ket{B_\gh}\,{\ket{B_\psi,\eta}}_{\rm R,NS}
  \,{\ket{B_\sgh,\eta}}_{\rm R,NS}~~,
\end{equation}
where
\begin{equation}
\label{bs5}
\ket{B_X} = \delta^{(d_\bot)}(\hat q - y)
\exp\biggl[-\sum_{n=1}^\infty \frac{1}{n}\,
\a_{-n}\cdot S\cdot
\tilde \a_{-n}\biggr]\,
\ket{0;k=0}~~,
\end{equation}
\begin{equation}
\label{bs6}
\ket{B_\gh} = \exp\biggl[\sum_{n=1}^\infty
(c_{-n}\tilde b_{-n}
 - b_{-n} \tilde c_{-n})\biggr]\,{c_0 +\tilde c_0\over 2}\,\ket{q=1}
\,\ket{\tilde q = 1}~~, \end{equation}
and, in the NS sector in the $(-1,-1)$ picture,
\begin{equation}
\label{bs7}
\ket{B_\psi,\eta}_{\rm NS} = \exp\biggl[\ii\eta\sum_{m=1/2}^\infty
\psi_{-m}\cdot S \cdot \tilde \psi_{-m}\biggr]
\,\ket{0}~~,
\end{equation}
\begin{equation}
\label{bs8}
\ket{B_\sgh,\eta}_{\rm NS} =
\exp\biggl[\ii\eta\sum_{m=1/2}^\infty(\gamma_{-m}
\tilde\beta_{-m} - \beta_{-m}
  \tilde\gamma_{-m})\biggr]\,
  \ket{P=-1}\,\ket{\tilde P=-1}~,
\end{equation}
or, in the R sector in the $(-1/2,-3/2)$ picture,
\begin{equation}
\label{bs9}
\ket{B_\psi,\eta}_\R = \exp\biggl[\ii\eta\sum_{m=1}^\infty
\psi_{-m}\cdot S \cdot \tilde \psi_{-m}\biggr]
\,\ket{B_\psi,\eta}_\R^{(0)}~~,
\end{equation}
\begin{equation}
\label{bs10}
\ket{B_\sgh,\eta}_\R =
\exp\biggl[ \ii\eta\sum_{m=1}^\infty(\gamma_{-m}
\tilde\beta_{-m} - \beta_{-m}
\tilde\gamma_{-m})\biggr]\,
 \ket{B_\sgh,\eta}_\R^{(0)}~~,
\end{equation}
where the superscript $^{(0)}$ denotes the zero-mode contribution
to be discussed momentarily.
The overall normalization factor $T_p$ can be unambiguously
fixed from the factorization of amplitudes of closed
strings emitted from a disk \cite{FRAU,cpb} and is the
tension of the D$p$-brane
\begin{equation}
\label{tens}
T_p = \sqrt{\pi} (2\pi\sqrt{\alpha'})^{3 - p}~~.
\end{equation}
To write explicitly the zero-mode parts
of the boundary state in the R-R sector, it is necessary to
introduce some further notation.
Let
\begin{equation}
\label{bs13}
\ket{A}\ket{\tilde B} = \lim_{z,\bar z\to 0} \, S^A(z)\tilde
S^B(\bar z) \, \ket{0}
\end{equation}
denote the vacuum for the fermionic zero-modes $\psi^\mu_0$ and 
${\tilde \psi}^\mu_0$, where $S^A$ and $\tilde S^B$ are the spin fields in
the 32-dimensional Majorana representation;
then, as shown in Ref. \cite{cpb}, we have
\beq
\label{bsr0}
\ket{B_\psi,\eta}_\R^{(0)} =
{\cal M}_{AB}^{(\eta)}\,\ket{A} \ket{\tilde B}~~, 
\eeq
where
\begin{equation}
\label{bs14}
{\cal M}^{(\eta)} = C\Gamma^0\Gamma^{l_1}\ldots
\Gamma^{l_p} \,\left(
\frac{1+\ii\eta\Gamma_{11}}{1+\ii\eta}\right)~~,
\end{equation}
with $C$ being the charge conjugation matrix and $l_i$ labeling the
space directions of the D-brane world volume.
Finally, if $\ket{P=-{1/ 2}}\,\ket{\tilde P=-{3/ 2}}$
denotes the superghost vacuum in the $(-1/2,-3/2)$ picture that is
annihilated by
$\beta_0$ and $\tilde \gamma_0$, we have \cite{YOST}
\beq
\label{bsrsg0}
\ket{B_\sgh,\eta}_\R^{(0)} =
\exp\left[\ii\eta\gamma_0\tilde\beta_0\right]\,
  \ket{P=-{1/ 2}}\,\ket{\tilde P=-{3/ 2}}~~.
\eeq
\par
Before using the boundary state to compute amplitudes involving
D-branes, one must perform the GSO projection.
In the NS-NS sector the projected state is
\begin{equation}
\label{bs22a}
\ket{B}_\NS  \equiv  {1 -(-1)^{F+G}\over 2}\,\, {1 -(-1)^{\tilde F+\tilde
G}\over 2}\, \ket{B,+}_\NS ~~,
\end{equation}
where $F$ and $G$ are the fermion and superghost number operators
\begin{equation}
{F} = {\sum_{m=1/2}^{\infty}\psi_{-m} \cdot \psi_m}
~~,~~
{G} = {- \sum_{m=1/2}^{\infty}
\left( \gamma_{-m}  \beta_m + \beta_{-m} \gamma_m
\right)} ~~.
\label{fergh}
\end{equation}
After some simple algebra, it is easy to see that
\begin{equation}
\label{bs22ab}
\ket{B}_\NS
= {1\over 2} \Big( \ket{B,+}_\NS - \ket{B,-}_\NS \Big)
\end{equation}
In the R-R sector the GSO projected boundary state is
\begin{equation}
\label{bs22ba}
\ket{B}_\R  \equiv  {1 +(-1)^{p} (-1)^{F+G}\over 2}\,\,
{1 - (-1)^{\tilde F+\tilde G}\over 2}\, \ket{B,+}_\R ~~.
\end{equation}
where $p$ is even for Type IIA and odd for Type IIB, and
\begin{equation}
(-1)^{F} = \Gamma_{11}(-1)^{\sum\limits_{m=1}^{\infty}\psi_{-m}
\cdot \psi_m}
~~,~~
{G} = - \gamma_0 \beta_0 - \sum\limits_{m=1}^{\infty}
\left[ \gamma_{-m} \beta_m + \beta_{-m}\gamma_m \right]~~.
\label{fermion}
\end{equation}
After some straightforward manipulations, one can check that
\beq
\label{bs22bb}
\ket{B}_\R  =
    {1\over 2} \Big( \ket{B,+}_\R + \ket{B,-}_\R\Big)~~.
\end{equation}
For later convenience we now rewrite the boundary state $\ket{B}_\R$
using 16-dimensional chiral and antichiral spinor indices $\a$ and
$\dot\a$ for Majorana-Weyl fermions. Then, for the Type IIA theory
we have
\bea
\ket{B}_\R &=& \frac{T_p}{2}
\ket{B_X}\, \ket{B_\gh}\,\left\{
\left(C\Gamma^0\Gamma^{l_1}\ldots\Gamma^{l_p}\right)_{\a\b}
\cos\left[\gamma_0\tilde\beta_0+
\Theta\right]
\ket{\a}_{-1/2}\ket{\tilde \b}_{-3/2}\right.
\nonumber \\
&&+\left.
\left(C\Gamma^0\Gamma^{l_1}\ldots\Gamma^{l_p}\right)_{\dot \a\dot \b}
\sin\left[\gamma_0\tilde\beta_0+
\Theta\right]
\ket{\dot\a}_{-1/2}\ket{\tilde{ \dot\b}}_{-3/2}\right\}~~,
\label{cossin}
\ena
where
\beq
\Theta=
\sum_{m=1}^\infty(\psi_{-m}\cdot S \cdot \tilde \psi_{-m}
+\gamma_{-m} \tilde\beta_{-m} - \beta_{-m}
\tilde\gamma_{-m})~~,
\label{theta}
\eeq
and we have abbreviated $\ket{\a,P=\ell}$ with $\ket{\a}_\ell$. Note that
one also has the following identity
\beq
\ket{\a}_\ell = \lim_{z\to 0} S^\a(z)\,{\rm e}^{\ell \phi(z)}\ket{0}~~.
\label{alphal}
\eeq
where $\phi$ is the chiral boson of the superghost
fermionization formulas
\beq
\gamma(z) = {\rm e}^{\phi(z)}\,\eta(z)~~~,~~~
\beta(z) = \partial\xi(z)\,{\rm e}^{-\phi(z)}~~.
\label{bosoniz}
\eeq
The boundary state for the Type IIB theory can be simply obtained
from \eq{cossin} by changing the chirality of the left moving fermions
according to \eq{bs22ba},
and explicitly reads 
\bea
\ket{B}_\R &=&\!\!  \frac{T_p}{2}
\ket{B_X}\, \ket{B_\gh}\,\left\{
\left(C\Gamma^0\Gamma^{l_1}\ldots\Gamma^{l_p}\right)_{\dot\a\b}
\cos\left[\gamma_0\tilde\beta_0+
\Theta\right]
\ket{\dot\a}_{-1/2}\ket{\tilde { \b}}_{-3/2}\right.
\nonumber \\
&&+\left.
\left(C\Gamma^0\Gamma^{l_1}\ldots\Gamma^{l_p}\right)_{ \a\dot\b}
\sin\left[\gamma_0\tilde\beta_0+
\Theta\right]
\ket{\a}_{-1/2}\ket{\tilde{\dot\b}}_{-3/2}\right\}~~,
\label{cossin1}
\ena
with $p$ odd.
\par
We would like to stress that the boundary states $\ket{B}_{\NS,\R}$
are written in a definite picture $(P,\tilde P)$
of the superghost system, where
\beq
P=\oint \frac{dz}{2\pi \ii }(-\partial\phi +\xi\, \eta)
\label{picture}
\eeq
and $\tilde P= -2 -P$ in order to soak up the anomaly in
the superghost number. In particular we have chosen $P=-1$ in the NS
sector and $P=-1/2$ in the R sector, even if other choices would have been
possible in principle \cite{YOST}. Since $P$ is half-integer in the R sector,
the boundary state $\ket{B}_{\R}$ has always $P\not=\tilde P$, and thus, as
suggested in Ref. \cite{sagnotti}, it
can couple only to R-R states in the asymmetric picture
$(P,\tilde P)$. However, the crucial point is to observe
that the massless R-R states in the $(-1/2,-3/2)$ picture may
contain a part that is proportional to the R-R {\it potentials},
as opposed to the standard massless R-R states in the
symmetric picture $(-1/2,-1/2)$ that are always proportional to the R-R field
strengths. This property, which has been already exploited in Ref. \cite{cpb}
to derive the D-brane effective action and the long-distance behavior
of the D-brane solutions from the boundary state formalism, will be used
later on to study the interaction between two D-branes. Many details
are contained in \secn{RRstate} and in the Appendices. Finally, we observe
that it is possible to fermionize the boundary states
$\ket{B}_{\NS,\R}$ using \eq{bosoniz}. However, as shown in Ref. \cite{YOST}
the equivalent boundary states in the $(\phi,\eta,\xi)$ system contain an
infinite number of terms
corresponding to the infinite possibilities of satisfying (\ref{picture}).
\par
We conclude this section by writing the conjugate boundary state
\begin{equation}
  \label{bs23}
 {}_{\rm R,NS}\bra{B,\eta} = {T_p\over 2}
    \,\bra{B_X}\,\bra{B_\gh}\,\,{}_{\rm R,NS}{\bra{B_\psi,\eta}}
  \,\,{}_{\rm R,NS}{\bra{B_\sgh,\eta}}
\end{equation}
which we will use in the calculation of the amplitudes between D-branes.
The explicit expressions for the various factors in \eq{bs23}
can be obtained by solving the overlap relations for the conjugate
boundary state and are given by the obvious counterparts of
Eqs. (\ref{bs5})-(\ref{bs10}). For example the fermion and the superghost
contributions are
\begin{equation}
  \label{bs26}
  {}_\NS\bra{B_\psi,\eta} = \bra{0}\,
  \exp\biggl[- \ii\eta\sum_{m=1/2}^\infty
  \psi_m\cdot S \cdot \tilde\psi_m\biggr]
\end{equation}
and
\begin{equation}
  \label{bs27}
  {}_\NS\bra{B_\sgh,\eta} = \bra{P=-1}\,\bra{\tilde P=-1}\,
  \exp\biggl[\ii\eta\sum_{m=1/2}^\infty(
  \beta_m \tilde\gamma_m - \gamma_m
  \tilde\beta_m)\biggr]
\end{equation}
in the NS-NS sector, and
\begin{equation}
  \label{bs28}
  {}_\R\bra{B_\psi,\eta} = \bra{A}\, \bra{\tilde B} \,
  {\cal N}_{AB}^{(\eta)}\,\,
  \exp\biggl[ -\ii\eta\sum_{m=1}^\infty
  \psi_m\cdot S \cdot \tilde\psi_m\biggr]~~,
\end{equation}
with
\beq
{\cal N}^{(\eta)} = {\Gamma^0}^T\,{\cal M^{(\eta)}}\,\Gamma^0
= (-1)^p\,C\Gamma^0\Gamma^{l_1} \ldots \Gamma^{l_p}\,\left(
\frac{1-\ii\eta\Gamma_{11}}{1+\ii\eta}\right)~~,
\label{matrix}
\eeq
and
\begin{equation}
  \label{bs29}
  {}_\R\bra{B_\sgh,\eta}=\bra{P=-{3/ 2}}\,\bra{\tilde P=-{1/2}}\,
  \exp\left[\ii\eta\beta_0\tilde\gamma_0\right]\,
  \exp\biggl[\ii\eta\sum_{m=1}^\infty(
  \beta_m \tilde\gamma_m -\gamma_m \tilde\beta_m )\biggr]
\end{equation}
in the R-R sector.
%\vskip 1cm
\sect{Interaction between a $p$ and a $p'$ brane}
\label{interaction}
%\vskip 0.5cm
In this section we study the static interaction between a D-brane
located at $y_1$, and a D-brane located at
$y_2$, with $NN$ directions common to the brane
world-volumes, $DD\geq 1$ directions transverse to both,
and $\nu = (10 -NN -DD)$ directions of mixed type.
We will not consider istantonic D-branes \cite{green},
hence also $NN\geq 1$.
The two D-branes simply interact via tree-level exchange of
closed strings whose propagator is
\beq
D= {\alpha'\over 4\pi} \int{d^2 z\over |z|^2} z^{L_0}\,
\bar z^{\tilde L_0}~~,
\label{prop}
\end{equation}
so that the static amplitude
is given by
\begin{equation}
  \label{bs32}
 {\cal A} = \bra{B^1}~D~\ket{B^2}~~,
\end{equation}
where $\ket{B^1}$ and $\ket{B^2}$ are the boundary states
describing the two D-branes\footnote{In \eq{bs32} it is
understood the usual insertion of the $(b,c)$ zero-modes
that is necessary to soak up the ghost anomaly of
the cylinder diagram (see for example Ref. \cite{callan}).}.
\par
The evaluation of ${\cal A}$ in the NS-NS sector presents
no difficulties and can be performed
starting from the definitions given in \secn{boundary}
and using standard string
techniques. Here we simply quote the final result, namely
\begin{eqnarray}
\label{ansns}
{\cal A}_{\rm{NS-NS}}
&=& \frac{V_{NN}}{2\pi} (8\pi^2\alpha')^{-{NN\over 2}}
\int_0^\infty
{dt} \left(\pi\over t \right)^{DD\over
2}\,{\rm e}^{-\Delta Y^2 /(2\alpha' t)}\,\nonumber\\
& & \times \left[ \left({f_3\over f_1}\right)^{8 -\nu}
  \left({f_4\over f_2}\right)^{\nu}
  - \left({f_4\over f_1}\right)^{8 -\nu}
  \left({f_3\over f_2}\right)^{\nu}\right]~~,
\end{eqnarray}
where $V_{NN}$ is the common world-volume of the two D-branes, $\Delta Y$
is the transverse distance between them, and the functions
$f_i$ are, as usual, given by
\begin{eqnarray}
  \label{fi}
  f_1=q^{{1\over 12}} \prod_{n=1}^\infty (1 - q^{2n}) ~~~~~~&,&~~
  f_2=\sqrt{2}q^{{1\over 12}} \prod_{n=1}^\infty (1 + q^{2n}) ~~,
  \nonumber\\
  f_3=q^{-{1\over 24}} \prod_{n=1}^\infty (1 + q^{2n -1})  ~~&,&~~
  f_4=q^{-{1\over 24}} \prod_{n=1}^\infty (1 - q^{2n -1}) ~~,
\end{eqnarray}
with $q={\rm e}^{-t}$. It is interesting to notice that
the two terms in the square brackets of \eq{ansns} come respectively
from the NS-NS$(-1)^{(F+G)}$ and the NS-NS sectors of the
exchanged closed string, which, under the
transformation $t\to{1}/{t}$, are mapped into the
NS and R sectors of the open string suspended between
the branes. Notice that ${\cal A}_{\rm{NS-NS}}=0$ if
$\nu=4$.
\par
On the contrary, the evaluation of ${\cal A}$ in the R-R sector
requires more care due to the presence of zero-modes in both
the fermionic matter fields and the bosonic superghosts.
To perform the calculation it is convenient to use boundary states
before the GSO projection, and thus consider
\beq
{\cal A}_{\rm{R-R}}(\eta_1,\eta_2) =~
 _{\rm{R}}\!\bra{B^1,\eta_1}~D~\ket{B^2,\eta_2}_{\rm{R}}~~.
\label{arr0}
\eeq
As before, it is not difficult to compute
the contribution of $X^\mu$, $(b,c)$ and also of the
non-zero-modes of $\psi^\mu$ and $(\beta,\gamma)$. Indeed,
after some algebra, we find
\bea
{\cal A}_{\rm{R-R}}(\eta_1,\eta_2) &=&
 \frac{V_{NN}}{\pi} (8\pi^2\alpha')^{-{NN\over 2}}
2^{-{\nu\over 2}}\int_0^\infty
{dt}
\left(\pi\over t \right)^{DD\over 2}\,
{\rm e}^{-\Delta Y^2 /(2\alpha' t)}\label{arr1}\\
& & \times \left[ 2^{\nu-4} \left(\frac{f_2}{f_1}\right)^{8-2\nu}
\delta_{\eta_1\eta_2,-1}+\delta_{\eta_1\eta_2,+1}\right]
~{}_\R^{(0)}\!\langle B^1,\eta_1 |
B^2,\eta_2\rangle_\R^{(0)}~~,
\nonumber
\ena
where
\beq
\ket{B,\eta}_\R^{(0)} =
\ket{B_\psi,\eta}_\R^{(0)}~\ket{B_\sgh,\eta}_\R^{(0)}
\label{bsr00}
\eeq
(see Eqs. (\ref{bsr0}) and (\ref{bsrsg0})).
Note that in \eq{arr1} it is essential {\it not} to separate the matter
and the superghost zero-modes.
In fact, a na{\"{\i}}ve evaluation of
${}_\R^{(0)}\!\langle B^1,\eta_1 |
B^2,\eta_2\rangle_\R^{(0)}$ would lead to a divergent or
ill defined result:
after expanding the exponentials in
${}_\R^{(0)}\!\langle B^1_\sgh,\eta_1 |
B^2_\sgh,\eta_2\rangle_\R^{(0)}$, all the infinite terms
with any superghost number contribute, and yield the divergent sum
$1+1+1+...$ if $\eta_1\eta_2=1$, or the alternating sum $1-1+1-...$ if
$\eta_1\eta_2=-1$. This problem has already
been addressed in Ref. \cite{YOST} and solved by introducing
a regularization scheme for the pure Neumann case ($NN=10$).
Here, we propose the
extension of this method to the most general case with D-branes.
\par
We define the scalar product in \eq{arr1} as follows
\beq
{}_\R^{(0)}\!\langle B^1,\eta_1 |
B^2,\eta_2\rangle_\R^{(0)} \equiv
\lim_{x\to 1} {}_\R^{(0)}\!\langle B^1,\eta_1 | \, {\cal R}(x) \,|
B^2,\eta_2\rangle_\R^{(0)} ~~,
\label{prodreg}
\eeq
where the regulator is
\beq
{\cal R}(x) = x^{2(F_0+G_0)}
\label{regulator}
\eeq
where $F_0$ and $G_0$ are the zero-mode parts of the
operators $F$ and $G$ (implicitly) defined 
in \eq{fermion}. 
After inserting the regulator, we can factorize the
scalar product and write
\bea
{}_\R^{(0)}\!\langle B^1,\eta_1 |
B^2,\eta_2\rangle_\R^{(0)}  &=& \lim_{x\to 1}~
\Bigg[~{}_\R^{(0)}\!\langle B^1_\psi,\eta_1 | \, x^{2F_0} \,
|B^2_\psi,\eta_2\rangle_\R^{(0)}
\nonumber \\
&&\times~{}_\R^{(0)}\!\langle B^1_\sgh,\eta_1 | \, x^{2G_0} 
\, |B^2_\sgh,\eta_2\rangle_\R^{(0)}~
\Bigg]~~.
\label{prodfact}
\ena
Let us first consider the superghost part. Remembering that
$G_0=-\gamma_0\b_0$, we 
simply have \bea
{}_\R^{(0)}\!\langle B^1_\sgh,\eta_1 | \, x^{2G_0} \,
|B^2_\sgh,\eta_2\rangle_\R^{(0)} &=&
\bra{-{3/2},-{1/2}}
{\rm  e}^{\ii\eta_1\beta_0\tilde\gamma_0}\,x^{-2\gamma_0\beta_0}\,
  {\rm e}^{\ii\eta_2\gamma_0\tilde\beta_0}\ket{-{1/2},-{3/2}}
\nonumber \\
&=&\frac{1}{1 - \eta_1 \eta_2 x^2}~~.
\label{bs50}
\ena
To discuss the regularization of the fermionic part, it is
convenient (even though not necessary) to perform a Wick
rotation $\Gamma^0\to \ii \Gamma^{10}$
and work with the $\Gamma$ matrices of ${\rm SO}(10)$.
Then, we can group them into five pairs, and for each pair
${\bf a} = (a_1,a_2)$, introduce fermionic annihilation and creation
operators
\begin{equation}
  \label{bs41}
  e^\pm_{\bf a} = {\Gamma^{a_1} \pm \ii \Gamma^{a_2}\over 2}~~,
\end{equation}
satisfying
\begin{equation}
  \label{bs42}
  \acomm{e^+_{\bf a}}{e^-_{\bf b}} = \delta_{{\bf a},{\bf b}}~~.
\end{equation}
The Hilbert space associated to each couple of $\Gamma$ matrices is
two-dimensional, and it is spanned by the states
$\ket{\uparrow}_{\bf  a}$ and $\ket{\downarrow}_{\bf  a} = e^-_{\bf a}
\ket{\uparrow}_{\bf  a}$, which are eigenvectors of the
number operator
\begin{equation}
  \label{bs43}
  N_{\bf a} = \comm{e^+_{\bf a}}{e^-_{\bf a}} =
  -\ii\, \Gamma^{a_1} \Gamma^{a_2}
\end{equation}
with eigenvalues $+1$ and $-1$ respectively.
Thus, the 32-dimensional space spanned by a ${\rm SO}(10)$
Majorana spinor is expressed as the direct product
of five copies of these two-dimensional spaces, each copy
for two different
directions. Note that in this way we explicitly break
${\rm SO}(10) \to {\rm SO}(2)^5$, but if all directions are equivalent,
the final results will still be ${\rm SO}(10)$ invariant
(see Ref. \cite{YOST}). In the case we are considering,
not all directions are on the same footing because of the
different boundary
conditions imposed by the D-branes.
Therefore, it is necessary to exert some care in pairing the
space directions
and the corresponding $\Gamma$ matrices if we want to have
meaningful
final results.
To this aim it is necessary that directions of mixed type be grouped
together, while the remaining NN or DD directions can be paired
as one wishes.  However, since for all
configurations we are considering, the time direction ({\it i.e.} the 
10$^{\rm th}$ after
Wick rotation) is always NN, and the 9$^{\rm th}$ direction can always
be chosen as DD, we group
them together, so that the prescription can
always be used. Note that
this choice amounts to specify the space direction in
which the light-cone
is oriented (namely the 9$^{\rm th}$), or
equivalently to mark in the
covariant formalism the direction which together with the time is
``canceled'' by the superghosts.

For the sake of simplicity, we now suppose that the NN directions of the
two D-branes are $(0,\ldots,p')$ and $(0,\ldots,p)$ with $p'\leq p$.
This specific choice allows us to perform explicitly the calculation, 
but it is not restrictive since any other configuration of D-branes can 
be reduced to it by a sequence of T-duality and parity transformations. 
%Using a series of T-duality transformations, we can always map
%our problem into an equivalent configuration of two D-branes whose NN
%directions are 
%$(0,\ldots,p')$  and $(0,\ldots,p)$, with $p'\leq p$. Thus, without any 
%loss of generality, in the following we will focus on this 
%case. 
Then, we group the ten $\Gamma$ matrices in five pairs, for example as
follows 
\bea
& \hbox{IIA} & ~\longrightarrow~~(1\,2)(3\,4)(5\,6)(7\,8)(9\,10)~,
\nonumber \\
& \hbox{IIB} & ~\longrightarrow~~(2\,3)(4\,5)(6\,7)(1\,8)(9\,10)~.
\ena
The next task is to find the expression for $F_0$ in this basis.
To do so, we observe that, in the ten dimensional euclidean 
space~\cite{YOST},
\beq
(-1)^{F_0} = \prod_{\mu=1}^{10} \Gamma^\mu  
= \ii\,\prod_{k=1}^5 N_k~~,
\label{f00}
\eeq
where $N_k \equiv N_{{\bf a}_k}$. Then,
since $\exp\left({\ii\,N_k\,{\pi}/{2}}\right) = \ii N_k$,
we can rewrite
\eq{f00} as follows
\beq
(-1)^{F_0} = \prod_{k=1}^5
\exp\left({\ii\,N_k\,\frac{\pi}{2}}\right)~~,
\label{f01}
\eeq
from which we can define
\begin{equation}
F_{\rm 0}=  \frac{1}{2} \sum_{k=1}^{5} N_k
~~.
\label{fermzer}
\end{equation}
Thus, the regulator for the fermionic zero-modes is
\begin{equation}
\label{bs45b}
x^{2F_{\rm 0}} = x^{~\sum\limits_{k=1}^5 N_{k} }~~.
\end{equation}
\par
We are now in the position of computing explicitly the contribution
of the fermionic zero-modes to the D-brane amplitude. Indeed, we have
\beq
{}_\R^{(0)}\!\langle B^1_\psi,\eta_1 | \, x^{2F_0} \,
|B^2_\psi,\eta_2\rangle_\R^{(0)} =
\tr \left( x^{2F_0}\,{\cal M}^{(\eta_2)}\,C^{-1}\,
{{\cal N}^{(\eta_1)}}^T\,C^{-1}\right)~~,
\label{overlap}
\eeq
where, to obtain the right hand side, we have used the inner product
\begin{equation}
\label{bs47}
\left (\bra{A}\bra{\tilde B}\right)\,
\left(\ket{D}\ket{\tilde E}\right) =
  - \bra{A}D\rangle ~\bra{\tilde B}\tilde E\rangle
  = - (C^{-1})^{AD}(C^{-1})^{BE}
\end{equation}
with the minus sign due to the exchange in the ordering of the
spinor states.
Inserting the explicit definitions (\ref{bs14}) and
(\ref{matrix}) of the matrices ${\cal M}$ and ${\cal N}$, after some
straightforward algebra, we find
\beq
{}_\R^{(0)}\!\langle B^1_\psi,\eta_1 | \, x^{2F_0} \,
|B^2_\psi,\eta_2\rangle_\R^{(0)} =
- \tr\!\!\left(\!x^{2F_0}
\prod_{\alpha} \Gamma^{\alpha}\!\right)\!
\delta_{\eta_1\eta_2,-1} - (-1)^p
\tr\!\!\left(\!x^{2F_0}\prod_{\alpha} 
\Gamma^{\alpha}~\Gamma_{11}\!\right)\!
\delta_{\eta_1\eta_2,+1}
\label{result0}
\eeq
where the index $\alpha$ runs over the $\nu$ directions of mixed type.
By introducing the fermionic
number operators as in \eq{bs43} and recalling that $\Gamma_{11}=
\prod_k N_k$, we can easily compute the
traces and get
\bea
\tr\left(x^{2F_0}\,\prod_{\alpha} \Gamma^{\alpha}\right) &=&
\prod_{k=1}^{\frac{10-\nu}{2}}
\tr\left(x^{N_k}\right)\,\prod_{l=1}^{\frac{\nu}{2}}
\tr\left(\ii\, x^{N_l}\,N_l\right)
\nonumber \\
&=&
\ii^{\frac{\nu}{2}}
\left(x+\frac{1}{x}\right)^{\frac{10-\nu}{2}}
\left(x-\frac{1}{x}\right)^{\frac{\nu}{2}}~~,
\label{result1} \\
\tr\left(x^{2F_0}\,\prod_{\alpha} \Gamma^{\alpha}~\Gamma_{11}\right) &=&
\prod_{l=1}^{\frac{\nu}{2}}
\tr\left(\ii\,x^{N_l}\right)\,\prod_{k=1}^{\frac{10-\nu}{2}}
\tr\left( x^{N_k}\,N_k\right)\nonumber \\
&=&
\ii^{\frac{\nu}{2}}
\left(x+\frac{1}{x}\right)^{\frac{\nu}{2}}
\left(x-\frac{1}{x}\right)^{\frac{10-\nu}{2}}~~.
\label{result2}
\ena
\par
Putting together the fermionic and superghost contributions as in
\eq{prodfact}, we finally obtain
\bea
{}_\R^{(0)}\!\langle B^1,\eta_1 |
B^2,\eta_2\rangle_\R^{(0)}  &=& \lim_{x\to 1}~
\left[-
\ii^{\frac{\nu}{2}}
\left(x+\frac{1}{x}\right)^{\frac{10-\nu}{2}}
\left(x-\frac{1}{x}\right)^{\frac{\nu}{2}}\,\frac{1}{1+x^2}\,
\delta_{\eta_1\eta_2,-1}
\right. \nonumber \\
&&\left. -(-1)^p\;\ii^{\frac{\nu}{2}}
\left(x+\frac{1}{x}\right)^{\frac{\nu}{2}}
\left(x-\frac{1}{x}\right)^{\frac{10-\nu}{2}}\,\frac{1}{1-x^2}\,
\delta_{\eta_1\eta_2,+1}
\right]
\nonumber \\
&=& -16 \,\delta_{\nu,0}\,\delta_{\eta_1\eta_2,-1} + 16 \,
\delta_{\nu,8}\,\delta_{\eta_1\eta_2,+1}~~.
\label{result3}
\ena
In our configuration, $\nu=8$ can be realized only for
$p'=0$ and $p=8$ so that we can understand the factor of $(-1)^p$; 
note that in this case there is a cancellation between the
zero coming from the fermionic zero-modes and the infinity coming
from the superghost zero-modes. This cancellation was qualitatively
suggested in the Appendix of Ref. \cite{GABE}, here we give a more 
rigorous derivation of it. It is interesting to observe that
\eq{result3} implies that
the R-R part of the D-brane amplitude vanishes unless
the two branes are parallel ($\nu=0$) or maximally orthogonal
($\nu=8$) to each other. Furthermore, the zero-mode contribution
to the R-R amplitude, which is the only one that survives in the field
theory limit, is numerically equal in both cases, and in particular it
is repulsive for $\nu=0$ and attractive for $\nu=8$.
This fact suggests that the nature of the force is
the same in the two cases and that the charges of
two D-branes
with $\nu=8$ can be essentially identified;
in \secn{factor} we will show that
this interpretation is indeed correct.
\par
We now write the final expression for the R-R amplitude. Inserting
\eq{result3} into \eq{arr1}, we get
\bea
{\cal A}_{\rm{R-R}}(\eta_1,\eta_2) &=&
 \frac{V_{NN}}{\pi} (8\pi^2\alpha')^{-{NN\over 2}}
\int_0^\infty {dt}
\left(\pi\over t \right)^{DD\over 2}\,
{\rm e}^{-\Delta Y^2 /(2\alpha' t)}
\nonumber\\
& & \times \left[ -\left(\frac{f_2}{f_1}\right)^{8}\,
\delta_{\nu,0}\,
\delta_{\eta_1\eta_2,-1}+\delta_{\nu,8}
\,\delta_{\eta_1\eta_2,+1}\right]~~,
\label{arr2}
\ena
from which we immediately obtain the GSO projected amplitude
\beq
{\cal A}_{\rm{R-R}}
= \frac{V_{NN}}{2\pi} (8\pi^2\alpha')^{-{NN\over 2}}
\int_0^\infty {dt}
\left(\pi\over t \right)^{DD\over 2}\,
{\rm e}^{-\Delta Y^2 /(2\alpha' t)}
~\left[ -\left(\frac{f_2}{f_1}\right)^{8}\,
\delta_{\nu,0}+\delta_{\nu,8}\right]~~.
\label{arr3}
\eeq
We conclude this section with a few observations.
First of all, the final result ${\cal A}_{\rm{R-R}}$
is Lorentz invariant, even if we used a non Lorentz
invariant regularization prescription to compute the
zero-mode contribution.
Secondly, the $\nu=0$ and $
\nu=8$ terms in \eq{arr3} come respectively from
the R-R$(-1)^{(F+G)}$ and the R-R sectors of the exchanged
closed string, which, under the transformation $t\to 1/t$,
are mapped into the ${\rm NS}(-1)^{(F+G)}$ and ${\rm R}(-1)^{(F+G)}$
sectors of the open string suspended between the branes.
Moreover, due to the ``abstruse identity'', the total D-brane amplitude
\beq
{\cal A}={\cal A}_{\rm{NS-NS}}+{\cal A}_{\rm{R-R}}
\label{total}
\eeq
vanishes if $\nu=0,4,8$; these are precisely the configurations of
two D-branes which break half of the supersymmetries of the Type II
theory and satisfy the BPS no-force condition.
\par
Note that for the $\nu=8$
systems the repulsive NS-NS force is canceled
by the attractive contribution of the R-R sector
that contains an insertion of $\Gamma_{11}$
(see Eqs. (\ref{result0}) and (\ref{result2})). Since
it is not parity invariant, this contribution becomes repulsive
if a parity transformation along the
DD direction ({\it i.e.} $X^9$) is performed.
As pointed
out in Ref. \cite{ohta}, such a parity transformation maps the
original $\nu=8$ system to a new configuration in which one of the
two D-branes has passed through the other and exchanged its
relative orientation. Therefore, after this exchange,
the total force between the D-branes becomes
\begin{eqnarray}
\label{aparity}
{\cal A}'
&=& \frac{V_{1}}{2\pi} (8\pi^2\alpha')^{-{1\over 2}}
\int_0^\infty
{dt} \left(\pi\over t \right)^{1\over
2}\,{\rm e}^{-\Delta Y^2 /(2\alpha' t)}\,\left[
  \left({f_4\over f_2}\right)^{8}
  - \left({f_3\over f_2}\right)^{8}-1\right]\nonumber\\
&=& \frac{V_1}{2\pi\a'} \,|\Delta Y|\,\, .
\end{eqnarray}
This force can be cancelled and the BPS condition restored only if
a fundamental string with tension $1/(2\pi\a')$ is created
when one D-brane passes through another in a $\nu=8$ system.
Notice that, according to this argument, this string
creation is an effect of the change in
the sign of the R-R
contribution due to a parity
transformation and {\it not} to a charge conjugation
which would transform one brane into an anti-brane
\cite{GABE}. As a matter of fact, in the case of maximally orthogonal 
branes, both transformations lead to the same result and
are equivalent because the parity preserving part of the R-R
sector vanishes. However, this does not happen in general when
the two D-branes are tilted with respect to each other
\cite{jabbari,ohta} since in this case, both the parity preserving
and the parity violating parts are non-zero. Finally, as
we have explicitly shown, the parity violating part of the R-R
force is non-vanishing only when there are $\nu=8$ directions
of mixed type; therefore, in all other
cases there is no fundamental string creation
when the D-branes pass through each other.
Another physical situation, where this parity violating
contribution is not trivial, is the magnetic interaction between
a D-brane and a dyon, recently examined in \cite{BIS}.
%\vskip 1cm
\sect{Asymmetric BRST invariant R-R states}
\label{RRstate}
%\vskip 0.5cm
Like in any string scattering amplitude, also for D-branes one can ask
which states mediate the interactions, {\it i.e.}
in our formalism, which states couple to the boundary state
and factorize the amplitude ${\cal A}$. To answer this question,
it is first necessary to discuss in some
detail the structure of the closed string states. Since
we are ultimately interested in the supergravity
interpretation of the D-brane amplitudes, we will
limit our considerations to the lowest mass level.
\par
In the NS-NS sector, where no zero-modes are present
and no particular technicalities are needed, the structure
of the massless states is simple. Nevertheless we find
useful to recall here their key properties, since these
will be shared also by the R-R states.
As shown in \secn{boundary}, the boundary state $\ket{B}_{\NS}$
is in the $(-1,-1)$ picture of the NS-NS sector, and
thus it can be directly saturated with
the usual symmetric states created by the vertex operators
\beq \label{VNS}
V_{\NS}(k;z,\bar z) =
\e_{\mu\nu} \,{V}_{-1}^\mu (k/2; z)
{\widetilde{V}}_{-1}^\nu (k/2; \bar z)~~,
\eeq
where
\beq \label{NS-1}
{V}_{-1}^\mu (k; z) = c(z)\,\psi^\mu (z)\,{\rm e}^{-\phi(z)} \;
{\rm e}^{ik\cdot X(z)}~~,
\eeq
and the form of $\e_{\mu\nu}$ depends on the particular NS-NS field
considered; for example a graviton requires a symmetric and traceless
polarization tensor.
\par
The vertex operator (\ref{VNS}) describes a physical NS-NS state only if
it is BRST invariant. To see what are the consequences of this
requirement, we split the BRST
charge into three parts: ${ Q} = { Q}_0 + {Q}_1 +
{Q}_2$ where
\bea
{Q}_0 &=& \oint \frac{dz}{2\pi \ii} ~c(z) \left[ T_{X} (z) + T_{\psi} (x)+
T_{\beta \gamma} (z) + \partial c(z)  b(z) \right]~~,
\label{Q12}
\\
\null \nonumber \\
{Q}_1 &=& \frac{1}{2} \oint \frac{dz}{2 \pi \ii} ~
{\rm e}^{\phi(z)}
\eta(z) \psi(z) \cdot \partial X(z)~~,\hskip 0.3cm
{Q}_2 = \frac{1}{4} \oint
\frac{dz}{2 \pi \ii} ~b(z) \eta(z) \partial \eta (z) {\rm e}^{2 \phi(z)}~~.
\nonumber
\ena
The first term contains a ghost part and the
energy-momentum tensor $T$
of the various fields, the second term
is proportional to the supersymmetry
current, while the last one contains only
ghost and superghost fields.
Then, it is not difficult to see that
\bea
&& \comm{{Q}_0 +\tilde{ Q}_0\,}{ V_\NS(k;z,\bar z)} \propto
k^2 \Big(\partial c(z)+\bar\partial\tilde{c}(z)\Big)
V_\NS(k;z,\bar z)~~,
\label{NSQ0}\\ &&
\comm{{Q}_1 +\tilde{Q}_1\,}{ V_\NS(k;z,\bar z)} = 0~~
\Leftrightarrow ~~ k^\mu\e_{\mu\nu} = 0~~,
\label{NSQ1}\\ &&
\comm{{Q}_2 +\tilde{Q}_2\,}{ V_\NS(k;z,\bar z)} = 0~~.
\label{NSQ2}
\ena
The vertex (\ref{VNS}) is therefore BRST invariant only if
$k^2=0$ and  $k\cdot\e = 0$. These two conditions
are really {\em independent} of each other: they originate from
two different terms of the BRST charge and can be imposed separately.
This means that we can continue off shell
the vertex operator (\ref{VNS}), maintaining Eqs.~(\ref{NSQ1})-(\ref{NSQ2})
and breaking the BRST invariance only in a soft way, because,
whenever $k^2 \not =0$,
the commutator with ${ Q}_0+{\tilde Q}_0$ remains proportional to
the original vertex.
\par
As is well known, perturbative string theory is consistent only
if it is BRST invariant. However, a propagating closed string
({\it i.e.} with $k^\mu\not=0$)
emitted from a disk or a D-brane, has to be off shell.
In fact, the Neumann boundary conditions along the
time direction impose $k_0=0$, and thus $k^2\not=0$.
We believe that, by breaking the BRST invariance in a
soft way, {\it i.e.} by allowing only a non-vanishing
commutator with $Q_0+{\tilde Q}_0$ as in \eq{NSQ0},
it is still possible to obtain meaningful results
in the field theory limit. Indeed, in
Ref. \cite{cpb}, we were able to
derive the correct
coupling and the correct large distance behavior of the various
massless fields of a D-brane solution
by simply saturating the boundary state with
off-shell closed string states of the lowest mass level.
\par
What we have called soft BRST breaking can be seen as a
variant of the off-shell extension proposed in Ref.~\cite{RamThor}.
The basic idea is that off-shell vertices satisfying
Eqs.~(\ref{NSQ0})-(\ref{NSQ2}) are representatives of
cohomology classes of the operator $Q'+{\tilde Q'}$ where
\beq
\label{q'}
Q'=Q-c_0 L_0 -b_0 M ~~~,~~~
M=- 2\left(\sum_{n>0} n c_{-n} c_n +\sum_{m\geq 0}
\gamma_{-m}\gamma_{m}\right) ~~,
\eeq
and similarly for the right moving part.
Since 
\beq
(Q'+{\tilde Q'})^2=-\frac{1}{2}\left[
(L_0+{\tilde L}_0)(M+{\tilde M}) + (L_0-{\tilde L}_0)(M-{\tilde M})
\right]~~,
\label{newq'}
\eeq
the operator $Q'+{\tilde Q'}$ can be seen as a BRST 
charge if it acts either on the usual on-shell states with 
$L_0={\tilde L}_0=0$, or on off-shell states such that
$L_0-{\tilde L}_0=0$ and $M+{\tilde M}=0$. The latter
conditions are precisely satisfied by the states we
are considering. Put differently, the propagating states 
emitted by D-branes or exchanged in their interactions
are off-shell representatives of the cohomology of the restricted
charge $Q'+{\tilde Q'}$. Indeed, the ghost zero-modes
that must be inserted in \eq{bs32} break the full BRST
invariance, but not the restricted one.
\par
Let us now turn to the R-R sector and, in particular, focus
our attention on the massless states of the type IIA theory
(of course our results are valid
also in the type IIB theory, as discussed in Appendix A).
The boundary state $\ket{B}_{\R}$ described in
\secn{boundary} is in the $(-1/2,-3/2)$ picture of the R-R
sector, and thus, to soak up the
superghost number anomaly, it can only couple to
states that are also in the asymmetric
$(-1/2,-3/2)$ picture. On the contrary, the R-R states
that are usually considered in the literature are
in the symmetric $(-1/2,-1/2)$ picture and are created by
the following vertex operators
\begin{equation}
V_\R(k;z,\bz) = \frac{1}{2\sqrt{2}}\,
(C{ F}^{\,(m+1)})_{\alpha{\dot{\beta}}}
\, {V}^\a_{-1/2}(k/2;z) \,
\tilde{V}^{\dot \b}_{-1/2}(k/2;\bz)~~,
\label{simver}
\end{equation}
where
\beq
(C{ F}^{\,(m+1)})_{\alpha{\dot{\beta}}} =
\frac{\left( C \Gamma^{{\mu_1} \dots
{\mu}_{m+1}} \right)_{\alpha{\dot{\beta}}}}{ (m+1)!} \,
F_{\mu_1 \dots \mu_{m+1}}
\label{simver1}
\eeq
with $m$ odd, and
\beq
{V}^\a_{\ell}(k;z)
= c(z)\,S^{\a}(z)\,{\rm
e}^{\ell\,\phi(z)} \,{\rm e}^{\ii k\cdot X(z)}~~.
\label{V-1/2}
\eeq
The form ${ F}^{(m+1)}$ has the right degree to be
interpreted as a R-R field strength of the type IIA theory; indeed
the vertex $V_\R$ is BRST invariant only if $k^2=0$, and
$d{ F}^{(m+1)}
=d\,*{ F}^{(m+1)}=0$ which are precisely the Bianchi and Maxwell
equations of a field strength.
\par
To explicitly construct the vertex operators in the
$(-1/2,-3/2)$ picture, we first require that they transform
into the symmetric vertices (\ref{simver}) with a picture
changing operation in the right
sector \cite{FMS}. The first guess is~\cite{sagnotti}
\begin{equation}
W^{(0)}(k;z,\bz) = (C{ A}^{\,(m)})_{\alpha{{\beta}}}
\, {V}^\a_{-1/2}(k/2;z) \,
\tilde{V}^{\b}_{-3/2}(k/2;\bz)~~,
\label{asim0}
\end{equation}
where ${ A}^{(m)}$ is given by
an expression similar to \eq{simver1}. Note that
the vertex (\ref{asim0}) satisfies
the same Type IIA GSO projection as
$V_\R$, and its
polarization contains a form ${ A}^{(m)}$ which has the
right degree to be identified with a R-R potential.
One can easily check that $W^{(0)}$ commutes with $Q_2$,
${\tilde Q}_2$ and ${\tilde Q}_1$, while the commutation
with $Q_1$ yields
\beq
\comm{{Q}_1 \,}{W^{(0)}(k;z,\bar z)} =\frac{1}{4\sqrt{2}}
\left(C\slash{k}{ A}^{(m)} \right)_{\dot\a \b}
\, \eta(z)\,{V}^{\dot\a}_{+1/2}(k/2;z)
\,\tilde{V}^{\b}_{-3/2}(k/2;\bz)~~.
\label{q1w0}
\eeq
The right hand side vanishes only if
$d{ A}^{(m)}= d\,*{ A}^{(m)}=0$. These conditions together
imply $k^2=0$, and thus guarantee that $W^{(0)}$ commutes also
with $Q_0+{\tilde Q}_0$. Thus, the vertex (\ref{asim0})
can be made BRST invariant, but then it
describes only a pure gauge potential. Moreover, since
the two linear conditions coming from
\eq{q1w0} imply $k^2=0$, the BRST invariance
is badly broken when $W^{(0)}$
is na{\"{\i}}vely extended off-shell.
\par
In order to describe non trivial gauge potentials in the
asymmetric picture of the R-R sector, in Ref. \cite{cpb}
we proposed a generalization of \eq{asim0}.
However, the vertex operators constructed there
contain explicitly the bare field $\xi$,
and thus do not belong to the $(\b ,\gamma)$ system, as it is
clear from the bosonization formulas (\ref{bosoniz}).
Another possibility
to describe potentials with non vanishing field strengths, is
given by the following vertex
operator\footnote{The overall factor of $1/2$ has been introduced 
for later convenience.}
\beq
W_G(k;z,\bz) = {1\over 2}\left[C\!\left(
{{A}}^{\,(m)} +
{{A}}^{\,(m+2)}\right) \right]_{{\alpha} {\beta}}
\, { V}^\a_{-1/2}(k/2;z) ~
\tilde{V}^{\b}_{-3/2}(k/2;\bz)~~.
\label{WG}
\eeq
Again, $W_G$ is invariant under
$Q_2$, ${\tilde Q}_2$ and ${\tilde Q}_1$,
while the commutation with the linear part of
the left BRST charge now yields
\beq
\comm{{Q}_1 \,}{W_G(k;z,\bar z)} =\frac{1}{8\sqrt{2}}
\left[C\slash{k}\left({ A}^{(m)} +
{ A}^{(m+2)}\right)\right]_{\dot\a \b}
\, \eta(z)\,{ V}^{\dot\a}_{+1/2}(k/2;z)\,
\tilde{V}^{\b}_{-3/2}(k/2;\bz)~~.
\label{1step}
\eeq
Using the properties of the $\Gamma$ matrices, one can see that
the right hand side is zero if
\beq
d\,*{ A}^{(m)} = 0 ~~,~~d{ A}^{(m)}+*\,d\,*{ A}^{(m+2)}=0
~~,~~d{ A}^{(m+2)}=0~~,
\label{scond}
\eeq
so that \eq{WG} describes a non trivial potential ${ A}^{(m)}$
in the Lorentz gauge and a pure gauge field ${ A}^{(m+2)}$
which decouples from all physical amplitudes.
Note that the conditions (\ref{scond}) together
imply that $k^2=0$, so that
$W_G$ commutes also with $Q_0+{\tilde Q}_0$. Finally,
by performing a picture changing in the right sector, one can recover
the symmetric vertex operator (\ref{simver}) with ${F}^{(m+1)}=
d{A}^{(m)}$. Hence, $W_G$ satisfies all the requirements
to be an acceptable R-R vertex operator of the closed string.
Just like $V_\R$, also the
asymmetric vertex (\ref{WG}) has definite
left and right superghost numbers, namely $G_0={\tilde G}_0=0$,
even though the only condition that is really
necessary in closed
string theory is $G_0+\tilde{G}_0=0$. On the other hand, contrarily to what
usually happens with the other vertex operators,
the polarization of $W_G$ is the sum of two different terms
corresponding to two different fields. Moreover, since the constraints
for the commutation of $W_G$ with $Q_1+{\tilde Q}_1$
imply the mass-shell condition $k^2=0$, again the BRST invariance is
badly broken when the vertex (\ref{WG}) is extended off-shell.
Thus, $W_G$ commutes with the restricted charge
$Q'+{\tilde Q'}$ only if $k^2=0$, and hence cannot 
represent the off-shell states that mediate the interactions
between D-branes.
\par
However, there exists yet another possibility to
write a non trivial vertex operator in the asymmetric picture.
In fact, let us consider again \eq{asim0} with $d{ A}^{(m)}\not=0$.
To cancel the right hand side of \eq{q1w0}, instead of
introducing a new gauge potential as in \eq{WG},
we add to
$W^{(0)}$ the following vertex
\begin{equation}
W^{(1)}(k;z,\bz) =- (C{ A}^{\,(m)})_{\dot\a \dot\b}
~\eta(z) {V}^{\dot\a}_{+1/2}(k/2;z) ~
\bar\partial{\tilde \xi}(\bz)
\tilde{V}^{\dot\b}_{-5/2}(k/2;\bz)~~, 
\label{asim1} 
\end{equation}
which commutes with $Q_2$ and ${\tilde Q}_2$,
but not with ${\tilde Q}_1$.
Indeed, one finds
\beq
\comm{{\tilde Q}_1 \,}{ W^{(1)}(k;z,\bar z)} =\frac{1}{4\sqrt{2}}
\left(C{ A}^{(m)}\slash{k} \right)_{\dot\a \b}
\, \eta(z)\,{V}^{\dot\a}_{+1/2}(k/2;z) \,
\tilde{V}^{\b}_{-3/2}(k/2;\bz)~~.
\label{q1w1}
\eeq
The right hand sides of Eqs. (\ref{q1w0}) and
(\ref{q1w1}) have the same structure and can compensate
each other if
\beq
\slash{k}{ A}^{(m)}+{ A}^{(m)}\slash{k} = 0 ~~,
\label{lorentz}
\eeq
that is, if the potential satisfies the Lorentz gauge condition
$d\,*{ A}^{(m)}=0$.
However, since the commutator of $Q_1$ with $W^{(1)}$ is not zero,
we must add another term $W^{(2)}$ to repair
the BRST invariance. The details of this iterative
construction are given in Appendix A, and the final result is a
vertex operator like
\beq
W(k;z,\bz) = \sum_{M=0}^\infty W^{(M)}(k;z,\bz)~~,
\label{Rvertex}
\eeq
where the infinite terms are recursively determined by asking
that the commutator of $Q_1$ with $W^{(M)}$ is
canceled by the commutator of ${\tilde Q}_1$ with
$W^{(M+1)}$  if \eq{lorentz} is satisfied.
This construction ensures that $W$ commutes with $Q_1+{\tilde Q}_1$
if just the Lorentz gauge is imposed
without requiring the mass-shell constraint $k^2=0$. The latter is only
needed to make $W$ invariant also under $Q_0$ and ${\tilde Q}_0$.
Thus, contrarily to $W_G$, the new vertex operator $W$ can be extended
off-shell by breaking the BRST invariance in a soft way. For
this reason, it can be saturated with
the boundary state to obtain the
correct coupling between a D-brane and a R-R potential.
This property is related to the fact that the off-shell
vertices $W$ are representatives of the cohomology
classes of the charge $Q'+{\tilde Q'}$ (see \eq{q'}), where
the superghosts zero-modes have not been singled out
contrarily to what suggested in Ref.~\cite{RamThor}.
\par
Even if the vertex operator (\ref{Rvertex}) looks
rather complicated, the state that it creates has, instead,
a rather simple expression when written in the $(\beta,
\gamma)$ system. In fact, as shown in detail in Appendix
A, we have
\bea
\ket{W} &=&
\lim_{z,\bz\to 0} W(k;z,\bz) \,\ket{0}
\label{-3/2state} \\
&=&\left(C{ A}^{(m)}\right)_{\a\b}
\cos\left(\gamma_0{\tilde \b}_0\right)
\ket{\a;k/2}_{-1/2}\,\ket{{\tilde \b};k/2}_{-3/2}
\nonumber\\
&&+\left(C{ A}^{(m)}\right)_{\dot\a \dot\b}
\sin\left(\gamma_0{\tilde \b}_0\right)
\ket{\dot\a;k/2}_{-1/2}\,\ket{{\tilde {\dot\b}};k/2}_{-3/2}
~~,
\nonumber
\ena
where we have introduced the notation
\beq
\ket{\alpha;k}_{\ell}
\equiv \lim_{z\to 0} {V}^\a_\ell(k;z)~\ket{0}~~.
\label{ket0}
\eeq
The state $\ket{W}$ is similar in form to the zero-mode part
of the boundary state $\ket{B}_\R$ (see \eq{cossin}), and like the latter,
is an eigenstate of the total superghost number $G_0+{\tilde G}_0$ with eigenvalue
zero, even though the left and right numbers are not separately 
well-defined. Asymmetric R-R states with this property have been
considered also in Ref.~\cite{NATHAN}.
Despite their very different structure, the two on-shell
states $\ket{W}$ of \eq{-3/2state} and
$\ket{W_G}$, created by the vertex operator
(\ref{WG}), describe the same physical content and correspond to
two different gauge choices for the R-R potentials.
More precisely, as we show in Appendix B,
$\ket{W}$ and $\ket{W_G}$ are
in the same BRST cohomology class,
that is\footnote{For this relation to hold,
it is crucial to have the factor of $1/2$
in front of the vertex operator (\ref{WG}) that creates
$\ket{W_G}$.}
\beq
\ket{W} = \ket{W_G} + \left(Q+{\tilde Q}\right)\ket{\Lambda}~~.
\label{equivRS}
\eeq
Note that the BRST equivalence (\ref{equivRS})
guarantees that $\ket{W}$ has a well-defined norm,
even if it is created by a vertex operator that contains
an infinite number of terms. In fact, as we discuss
at length in the next section, there exists a well-defined scalar
product for the states $\ket{W}$ which utilizes the same regulator
${\cal R}(x)$ appearing in the scalar product of boundary states.
Finally, we would like to comment that the asymmetric
$(-1/2,-3/2)$ picture of the R-R sector does not necessarily require
that the states are proportional to the R-R gauge potentials. For
example, the state \cite{NATHAN}
\beq
\left(C{ F}^{(m+1)}\right)_{\a \dot \b}
(c_0+{\tilde c}_0) \,{\tilde \b}_0\,\ket{\alpha;k/2}_{-1/2}
\ket{\tilde {\dot \b}; k/2}_{-3/2}
\label{fstate}
\eeq
is BRST invariant if ${ F}^{(m+1)}$ satisfies the Maxwell equations
for a field strength, and transforms to a symmetric state in the
$(-1/2,-1/2)$ picture with a picture changing operation in the right
sector.  In this transformation only the algebraic
part of the BRST charge, ({\em i.e.} ${\tilde Q}_2$), plays a non
trivial role, contrarily to what happens with the states $\ket{W}$,
where it is ${\tilde Q}_1$ that acts non trivially and transforms
the potential into a field strength. However, states like
(\ref{fstate}) do not play any role in our discussion since,
having a non zero total superghost number,
they cannot couple to the boundary state.
\par
In the following section we demonstrate that the states $\ket{W}$
correctly factorize the R-R
amplitude between two D-branes, and use this result
to explain the Coulomb like
interaction of the $\nu=8$ systems  from a field theory point of view.
%\vskip 1cm
\sect{Factorization and supergravity analysis of the $\nu=8$ systems}
\label{factor}
%\vskip 0.5cm
{}From the point of view of supergravity the results
of \secn{interaction} may appear surprising: in fact, while
the absence of force between two parallel D-branes ($\nu=0$) can be
seen as the result of the cancellation among three Feynman
diagrams ({\em i.e.} the exchange of a graviton, a dilaton
and a $(p+1)$--form potential), it does not seem possible
to give a similar interpretation to the no-force condition
found for the $\nu=8$ configurations. In this case
the string calculation tells us that the total force due to the
exchange of NS-NS states is repulsive and independent
of the distance between the two D-branes; on the other hand,
even if the two D-branes carry R-R
charges of different type, there is a non-vanishing contribution
from the R-R sector that exactly cancels the repulsive NS-NS force.
In this section we show that there is a very simple
interpretation of such a R-R force also from the field
theory point of view. For definiteness
we will consider in detail the system of one D0 and one
D8-brane, but of course our conclusions are valid as well
in all other $\nu=8$ systems related to it by T-duality.
The crucial point is the following:
for the $\nu=8$
systems, the charges of the two D-branes are essentially
identified, produce the same R-R potential, and
therefore, just like in the $\nu=0$ systems,
the R-R interaction is simply due to the usual
Coulomb-like force between the D-branes. It should not be
surprising
that forms of different order can be identified.
One well-known example of this phenomenon is the
electro-magnetic duality of the Type II theories.
In fact, let us consider the massless R-R state
created by the symmetric vertex operator (\ref{simver})
\begin{equation}
|V\rangle =
 \frac{\left( C \Gamma^{{\mu_1} \dots
{\mu}_{m+1}} \right)_{\alpha{\dot{\beta}}}}{ 2\sqrt{2}(m+1)!} \,
F_{\mu_1 \dots \mu_{m+1}}
\,\ket{\alpha;k/2}_{-1/2}\,
\ket{\tilde{\dot{\beta}};k/2}_{-1/2}~~.
\label{usver}
\end{equation}
Due to the structure
of the polarization factor in \eq{usver}, it is easy
to see that the field strengths satisfy
the duality condition ${F}^{\,(m+1)} \simeq
*{ F}^{\,(9-m)}$.
Such a relation
imposes a duality constraint
also on the physical degrees of freedom of the
potentials ${ A}^{\,(m)}$. For instance,
if the space momentum of the state (\ref{usver})
lies entirely along the $9^{\rm th}$
direction, the duality condition for a 2-form reads
\beq
F_{01}=k_0 A_1 = -F_{2\ldots 9} = k_9 A_{2\ldots 8}~~.
\label{duality}
\eeq
Since the state (\ref{usver}) is massless,
one has also
$k_0= k_9$, and thus
\beq
A_1=A_{2\ldots 8}~~.
\label{duality0}
\eeq
In this way one recovers the
usual duality relation
for the potential fields ${ A}^{\,(m)} \simeq *{A}^{\,(8-m)}$
in the transverse space.
It is important to realize that
this relation involves only the physical transverse
degrees of freedom of the potentials, and does not hold for the non-physical
polarizations that are the analogues of the
scalar and longitudinal photons of electrodynamics.
Usually, such states are never taken into account in string theory
because they decouple from any physical amplitude.
However, as we shall see,
they are particularly relevant to the present discussion because a
mixture of longitudinal and scalar states
provides a local description
of the Coulomb force between D-branes,
which may appear
as an instantaneous effective interaction when just
the physical degrees of freedom are quantized.
Note that, if
only closed strings are present,
there are no R-R charged objects, and thus the
longitudinal and scalar polarizations do not appear as
propagating states either.
Even if these degrees of freedom always decouple,
we can still write some
states that describe them in a pure closed string 
framework. These are created by the vertex
operator (\ref{asim0}), and are explicitly
given by
\begin{equation}
\ket{W^{(0)}} = \frac{1}{m!}
\left(C\Gamma^{\mu_1 \dots \mu_{m}} \right)_{\alpha \beta}
A_{\mu_1 \dots \mu_m}
\ket{\alpha;k/2}_{-1/2}\,
\ket{\tilde{\beta};k/2}_{-3/2}~~.
\label{-3/2state1}
\end{equation}
In fact, as we discussed in \secn{RRstate}, $\ket{W^{(0)}}$
is BRST
invariant if $k^2=0$ and
$dA^{(m)}=d\,*A^{(m)}=0$ which are precisely
the conditions fulfilled by the mixture of
longitudinal and scalar polarizations that describes the Coulomb
interaction.
For example a 1-form potential of this type with $k_0=k_9$
is characterized by
\beq
A_0=A_9~~,~\hbox{and}~~A_i=0~~~\hbox{for}~~i=1,\ldots,8~~.
\label{unphypol}
\eeq
The spinorial structure of \eq{-3/2state1} implies
that the unphysical polarizations satisfy a 10-dimensional
Hodge duality similar to the one of the field strength; for instance,
in the case of a 1-form we have $A_9=-A_{01\ldots8}$, which combined with
\eq{unphypol} leads to the following relation
\beq
A_0 = - A_{01\ldots 8}~~.
\label{ident1}
\eeq
The two dualities
(\ref{duality0}) and (\ref{ident1}), and their obvious generalizations,
can be unified by saying that
the components of two potentials are identified if
they have the same longitudinal indices and their
{\em transverse} indices are complementary.
The unusual relation (\ref{ident1}) is of
no relevance in perturbative string theory where
the unphysical degrees of freedom always decouple, but it
becomes important when dealing with boundary states. In this case
it has remarkable
consequences: in fact, it implies that the charge
felt by $A_0$ is opposite to the charge felt by
$A_{01\ldots 8}$, and thus the attractive Coulomb
R-R force between a D0 and a D8 brane can be interpreted as
due to the exchange of longitudinal and scalar
polarizations identified according to \eq{ident1}. Note that this argument
is consistent with the results of
\secn{interaction}, where we have found
that, in $\nu=8$ systems, the R-R
force does not receive any correction from the massive string
states and is exactly the opposite of the force due
to the exchange of massless R-R
fields between two parallel D-branes.
To see that this interpretation is correct, we actually need to
verify that the states propagating between two D-branes satisfy
duality relations like (\ref{ident1}). Since these states are off-shell,
they can not be like $\ket{W^{(0)}}$ of \eq{-3/2state1}, but instead
are like $\ket{W}$ of \eq{-3/2state}.
Thus, henceforth we focus on asymmetric states of
this type and study the Hilbert space they generate.
As a first step, we define the conjugated state $\bra{W}$ which
satisfies the same GSO projection of $\ket{W}$, namely
\bea
\bra{W} &\! = \! &
{}_{-3/2}\bra{\dot\alpha;k/2}~
{}_{-1/2}\bra{\tilde{\dot\beta};k/2}
\,\cos(\beta_0\tilde{\gamma_0})\;(C A^{(m)})_{\dot{\alpha}
\dot{\beta}} \nonumber
\\ & &+~
{}_{-3/2}\bra{\alpha;k/2}~
{}_{-1/2}\bra{\tilde{\beta};k/2}
\,\sin(\beta_0\tilde{\gamma_0})\;
(C A^{(m)})_{{\alpha} {\beta}}~~.
\label{-3/2bra}
\ena
However, with this definition the na{\"{\i}}ve scalar product between
a bra and a ket is divergent or
ill defined due to the infinite contributions of the superghosts, just
like the na{\"{\i}}ve scalar product between two
boundary states. We overcome this problem by regularizing the
scalar product with the same prescription
used in \secn{interaction} for the
amplitude between two boundary states. Thus, we define
\beq
\langle W'\,,\,W\rangle \equiv
\lim_{x\to 1}\;\bra{W'}\,{\cal R}(x)\,\ket{W}~~,
\label{scalprod}
\eeq
where the regulator ${\cal R}(x)$ is given in \eq{regulator}.
It is now
easy to see that, with this prescription, the states
of \eq{-3/2state} have a definite
norm. For example, following the same procedure outlined in
\secn{interaction}, for the case of 1-forms we have
\bea
\langle W\,\,,W\rangle & = &
\lim_{x\to 1}~
\Bigg[{1\over 2(1+x^2)} \,
\langle A| \bra{\tilde B}
(C A^{(1)})_{AB}\,x^{2F_0}\,
(C A^{(1)})_{CD}
|C\rangle |\tilde D\rangle
\nonumber \\ && +
{1\over2(1-x^2)} \,
\langle A| \bra{\tilde B}
(C A^{(1)})_{AB}\, x^{2F_0}\,
(C A^{(1)}\Gamma_{11})_{CD}
|C\rangle |\tilde D\rangle
\Bigg]
\nonumber \\ & = &
8\, A_\mu A^\mu~~.
\label{1norm}
\ena
Note that the second line does not contribute, since the result of the
scalar product over the fermionic zero-modes
goes to zero faster than $(1-x^2)$ when $x\to
1$. The factor of 8 in the final result correctly counts the physical
degeneracy of the R-R vacuum, and is the product of the
superghost contribution ({\it i.e.} $1/2$) and of the chiral trace over the
$\Gamma$ matrices ({\it i.e.} $16$).
Thus, our regularization prescription makes
manifest the role of the superghost zero-modes in the R-R sector, that
is to halve the degeneracy of the fermionic vacuum. Moreover, with the
definition (\ref{scalprod}), the one-to-one correspondence
between $\ket{W}$ and $\ket{W_G}$ displayed in \eq{equivRS}
becomes an
isometry: in fact, 
the scalar product (\ref{scalprod}) reduces to the usual one when
$\ket{W}$ is written in terms of $\ket{W_G}$.
But the most striking feature of the scalar product (\ref{scalprod})
is that forms of different order are, in general,
not orthogonal to each other. To see this explicitly, let us
consider for example the subspace generated by
\bea
\ket{W_1} &=& (C\Gamma^1)_{\a\b} \cos(\gamma_0\tilde{\beta_0})\,
\ket{\a;k/2}_{-1/2}~\ket{\tilde \b;k/2}_{-3/2}
\nonumber \\
&&+~
(C\Gamma^1)_{\dot \a\dot \b} \sin(\gamma_0\tilde{\beta_0})\,
\ket{\dot \a;k/2}_{-1/2}~\ket{\tilde{\dot \b};k/2}_{-3/2}
\label{w1}~~,
\ena
and
\bea
\ket{W_{2\ldots8}} &=& (C\Gamma^{2\ldots8})_{\a\b}
\cos(\gamma_0\tilde{\beta_0})\,
\ket{\a;k/2}_{-1/2}~\ket{\tilde \b;k/2}_{-3/2}
\nonumber \\
&&+~
(C\Gamma^{2\ldots8})_{\dot \a\dot \b} \sin(\gamma_0\tilde{\beta_0})
\ket{\dot \a;k/2}_{-1/2}~\ket{\tilde{\dot \b};k/2}_{-3/2}
\label{w7}~~,
\ena
where the momentum lies only in the $0^{\rm th}$
and $9^{\rm th}$ directions
in order to keep the transversality condition.
These two states are not perpendicular to each other, because in the mixed
scalar product the analogue of the second line of \eq{1norm} gives a
non-vanishing contribution leading to
$\langle{W_1}\,,\,W_{2\ldots8}\rangle =8$.
We then conclude that, in the subspace spanned by $\ket{W_1}$ and
$\ket{W_{2\ldots8}}$, the scalar product (\ref{scalprod})
defines a degenerate metric proportional to
\beq
\left(\begin{array}{cc}
 1 & 1 \\ 1 &  1
\end{array}\right)
\label{metric}
\eeq
with a null-state
\beq
\ket{\zeta}=\ket{W_1}-\ket{W_{2\ldots8}}~~,
\label{null1}
\eeq
which decouples from all
amplitudes, even if boundary states are present. Thus, we can set
$\ket{\zeta}=0$, and identify the components of a R-R field along $\ket{W_1}$
and $\ket{W_{2\ldots8}}$, recovering in this way the off-shell
extension of the duality relation
(\ref{duality0}).
These arguments hold for all states with transverse polarizations,
while for longitudinal and scalar states the results are
slightly different, even if the analysis is similar. Let us consider,
for example, the 1-form state that carries the Coulomb interaction
(\ref{unphypol}), and continues it off-shell in the kinematic region
that is relevant for the study of D-brane interactions, namely
\bea
\ket{W_0} &=& (C\Gamma^0)_{\a\b} \cos(\gamma_0\tilde{\beta_0})\,
\ket{\a;k/2}_{-1/2}~\ket{\tilde \b;k/2}_{-3/2}
\nonumber \\
&&+~
(C\Gamma^0)_{\dot \a\dot \b} \sin(\gamma_0\tilde{\beta_0})\,
\ket{\dot \a;k/2}_{-1/2}~\ket{\tilde{\dot \b};k/2}_{-3/2}
\label{w0l}~~,
\ena
where $k_i=0$ for $i=0,\ldots, 8$. It is very easy to verify that
the following 9-form state
\bea
\ket{W_{0\dots8}} &=& (C\Gamma^{0\ldots8})_{\a\b} \cos(\gamma_0\tilde{\beta_0})\,
\ket{\a;k/2}_{-1/2}~\ket{\tilde \b;k/2}_{-3/2}
\nonumber \\
&&+~
(C\Gamma^{0\ldots8})_{\dot \a\dot \b} \sin(\gamma_0\tilde{\beta_0})\,
\ket{\dot \a;k/2}_{-1/2}~\ket{\tilde{\dot \b};k/2}_{-3/2}
\label{w9}
\ena
is not orthogonal to $\ket{W_0}$.
In fact, in the scalar product $\langle W_{0}\,,W_{0\ldots 8}\rangle$,
the two $\Gamma_0$
matrices present in both states cancel, and so the calculation becomes
identical to the one outlined for $\langle W_{1}\,,W_{2\ldots8}\rangle$.
Again, in the subspace
spanned by the states (\ref{w0l}) and (\ref{w9}), the scalar product
leads to a degenerate
metric proportional to
\beq
\left(\begin{array}{cc}
 ~~1 & -1 \\ -1 & ~~1
\end{array}\right)
\label{metric1}
\eeq
which admits the null state
\beq
\ket{\chi}=\ket{W_0}+\ket{W_{0\ldots8}}~~.
\label{null2}
\eeq
While on-shell the scalar and longitudinal polarizations like
(\ref{unphypol})
always decouple from physical
amplitudes, the presence of a D-brane forces an off-shell
continuation and only linear combinations like (\ref{null2})
decouple. When we set $\ket{\chi}=0$, we are led to identify
$A_0$ with $-A_{0\ldots8}$, recovering in this way  the duality
relation (\ref{ident1}) also off-shell.
Thus, the Hilbert space structure is responsible both for the
identification (\ref{duality0}), that is usually seen as an effect of
the GSO projection of the symmetrical states, and for the
identification (\ref{ident1}), that is proper only of the
longitudinal asymmetric states
(\ref{-3/2state}), since these have no symmetric counterparts.
\par
At the beginning of this section, we suggested that the linear
R-R potential in the $\nu=8$ systems is simply the analogue of the usual
Coulomb electric potential between two charges of opposite sign;
this is because the two D-branes produce essentially the same R-R field,
even if their charges seem different.
Now we can explicitly prove that this
interpretation is correct by identifying the R-R state exchanged in the
interaction between the two D-branes. In order to
exploit the explicit formulas that we have derived before,
we consider a system
of one 0-brane and one 8-brane, but it is clear
that this analysis is general and can be
applied to all $\nu = 8$ configurations. Focusing on the massless
sector, the R-R amplitude between
the two boundary states, ${\cal A}_{\R-\R}^0$,
can be factorized by writing the identity
operator with the asymmetric states of \eq{-3/2state}. In particular
the relevant terms are
\beq
\one ~=~ \ket{W_0}\,\bra{W_0} ~+~ P(\ket{\chi}) + \ldots~~,
\label{identity}
\eeq
where $P$ is a projector onto the one dimensional space generated by
the vector $\ket{\chi}$. From our previous analysis of the R-R Hilbert
space structure, we know that the combination $\ket{\chi}$ can
be ignored because it always decouples, while the 1-form state $\ket{W_0}$
has a non-zero overlapping with both the 0-brane and the 8-brane.
Inserting twice the identity (\ref{identity}) in the \eq{bs32}, it is
easy to see that the massless R-R contribution to the D-brane
interaction can be written as
\beq
{\cal A}_{\R-\R}^0=
{}_{R}\bra{B_{0}}~W_0\rangle~
\langle W_0 | D \ket{W_0}~
\langle W_0 \ket{B_{8}}_{R}~~,
\label{fact}
\eeq
where, henceforth, the state $\ket{W_0}$ is normalized
to one\footnote{This simply amounts to multiply \eq{w0l} by
$1/(2{\sqrt{2}})$.},
while all the other R-R states that give no contribution are understood.
With this normalization one can verify that, in the field theory limit,
the scattering amplitude among the asymmetric R-R states
and the NS-NS ones are correctly reproduced by
the following action
\beq
S = \int d^{10}x\,\sqrt{-g}\left[-\frac{1}{2(m+1)!}~
{\rm e}^{\frac{5-(m+1)}{\sqrt{2}}\kappa\phi}
\left( F^{(m+1)}_{\mu_1\ldots\mu_{m+1}}\right)^2 \right] ~~,
\label{RRact}
\eeq
where, as usual, $F^{(m+1)}=dA^{(m)}$.

We can now reinterpret the result of the factorization (\ref{fact})
as the missing Feynman diagram describing the R-R interaction
from a field theory point of view.
In fact, the closed string propagator $D$ in \eq{fact} simply
becomes the usual field theory propagator in the Feynman gauge $-1/k^2$,
that is consistent with the action (\ref{RRact}).
On the other hand, from Ref.~\cite{cpb}, we know that, when a boundary state
is directly saturated with a closed string state, one gets the field
theory vertices that are usually derived from the Born-Infeld lagrangian,
namely
\beq
\langle W_0| B_{0} \rangle_{R} = -\langle W_0| B_{8} \rangle_{R}
= -\sqrt{2}T_0\,V_{1}~~.
\label{coupling}
\eeq
The coupling of a D8-brane and a 1-form potential is not manifest at
the level of the D-brane action, because it is an effect of the duality
relation (\ref{ident1}). As is well known for the
type IIB theory and other similar cases, it is rather
difficult to implement a duality relation directly in
an action, and normally infinite auxiliary fields
are needed for this purpose.
%%%%%%%%%%%%%%%%%%%%%%%%%%%%%%%%
\vskip 2cm
\appendix{\Large {\bf {Appendix A}}}
\label{appa1}
\vskip 0.5cm
\renewcommand{\theequation}{A.\arabic{equation}}
\setcounter{equation}{0}
\noindent
In this appendix we prove the BRST invariance of the vertex
operator
\begin{equation}
\label{maa2}
W(k;z,\bar z) = \sum_{M=0}^\infty W^{(M)}(k;z,\bar z)
\end{equation}
introduced in \secn{RRstate} to
describe the emission of a R-R field in the asymmetric
$(-{1/2},-{3/2})$ picture.
We treat simultaneously Type IIA and Type IIB theories
assuming that the GSO projection is defined in \eq{bs22ba}
(with $p$ even for Type IIA, and $p$ odd for Type IIB), and use
32-dimensional spinor indices. For notational convenience, we introduce
also the chiral projectors
$\Pi_{q}\equiv (1 + (-1)^q\Gamma_{11})/2$, and the following
combinations:
\begin{eqnarray}
 {\cal V}^A_{-1/2 +M}(z) & = &
 {\partial}^{M-1} \eta (z) \dots \eta(z) c(z) S^A(z) {\rm e}^{(-{1\over 2}
 +M)\phi(z)}{\rm e}^{\ii k \cdot X(z)/2}~,\nonumber\\
 {\tilde{\cal V}}^A_{-3/2 -M}(\bar z) & = &
 \bar\partial^{M} \tilde\xi(\bar z) \dots \bar\partial 
\tilde\xi(\bar z)
 \tilde c(\bar z) \tilde S^A(\bz) {\rm e}^{(-{3\over 2}-M)
\tilde\phi(\bz)}
 {\rm e}^{\ii k \cdot \tilde
 X(\bar z) /2}~,
 \label{genefo3}
\end{eqnarray}
where $S^A$ are the spin fields and
$\phi$, $\xi$ and $\eta$ come from the superghost fermionization
(see \eq{bosoniz}).
With these notations and dropping for simplicity
the dependence on the momentum $k$, all terms in \eq{maa2}
can be written in a compact
way as
\begin{equation}
 \label{c10}
 W^{(M)}(z,\bar z) =
 a_M \,\left[\Pi_{p+M} C{ A}^{(m)}
 \Pi_{M}\right]_{AB}\, {\cal V}^A_{-1/2 + M}(z)\,
 \tilde {\cal V}^B_{-3/2 - M}(\bz)~~,
\end{equation}
where $a_0=1$ and the other coefficients $a_M$ will be specified later.
Converting to the Majorana-Weyl notation, one can easily recognize
for instance that \eq{c10} for $M=0$ in a Type IIA theory 
($p$ even and $m$ odd) reads
\begin{equation}
 \label{din1}
 \left(C{ A}^{(m)}\right)_{\alpha\beta} c(z) S^\alpha(z) {\rm e}^{-\phi(z)/ 2}
{\rm e}^{\ii k\cdot X(z)/2} ~
\tilde c(\bar z)\tilde S^\beta(\bar z) {\rm e}^{-3\tilde\phi(\bar z)/2}
{\rm e}^{\ii k\cdot\tilde X(\bz)/2}~~,
\end{equation}
which is precisely \eq{asim0}. Similarly, one can check that \eq{c10}
for $M=1$ reduces to \eq{asim1} if $a_1=1$.
We now study the BRST properties of the vertex $W$, and to do this we split
the BRST charge according to \eq{Q12}.
It is well known that any operator with conformal dimension $0$
and of the type $c(z) U(z)$, where
$U(z)$ is a primary field made up only by matter and
superghost fields, commutes with
${Q}_0$. Then, it follows that the commutation with $Q_0 +\tilde Q_0$ is
equivalent to the on-shell condition
\begin{equation}
 \label{maa1}
 \comm{Q_0\,}{W^{(M)}(z,\bz)} = \comm{\tilde Q_0\,}{W^{(M)}(z,\bz)} = 0 \,
 \Leftrightarrow k^2 = 0~~.
\end{equation}
It is also easy to show, by direct computation of the relevant
OPEs, that $W^{(M)}$ commutes with ${Q}_2$ and
$\tilde{Q}_2$.
The only non-trivial commutators are the ones with ${Q}_1$ and
$\tilde{Q}_1$. To evaluate them, we make use of the following OPEs \cite{Kost}
\begin{equation}
 \label{c7}
 (\psi^\mu)_1(w)\, S^A_{-n/2}(z) ~\sim ~{\ii\over \sqrt{2}}
 (w - z)^{n - 1\over 2}\, \left(\Gamma_{11}\Gamma^\mu\right)^A_{\, B}
 S^B_{-n/2 + 1}(z)~~,
\end{equation}
where we abbreviated $(\psi_{\mu} )_1 (z) \equiv
\psi_{\mu} (z) {\rm e}^{\phi (z)}$ and $S^A_{-n/2}(z)\equiv
S^A(z){\rm e}^{-{n\over 2}\phi(z)}$, and
\begin{equation}
\partial X^{\mu} (w)\, {\rm e}^{\ii k \cdot X(z)/2} ~\sim~
\frac{ \ii \left(k^{\mu}/2 \right) {\rm e}^{i k \cdot X(z)/2}}{w-z}~~,
\label{contra3}
\end{equation}
plus of course their tilded counterparts.
Straightforwardly enough one obtains
\begin{eqnarray}
\comm{{ Q}_1\,}{W^{(M)}(z,\bz)} & = &
{a_M (-1)^{p + M}\over4 \sqrt{2} M! }\,
\left(\Pi_{p+M+1} C \slash k {A}^{(m)} \Pi_{M}\right)_{AB}
\nonumber \\
&\times& {\cal
V}^A_{-{1\over 2} + (M+1)}(z)~\tilde{\cal V}^B_{-{3\over 2} - M}(\bz)
~~,\label{brstM} \\
\comm{\tilde{Q}_1}{W^{(M+1)}(z,\bz)} & = &
{-a_{M+1} (M+1)!\over 4 \sqrt{2}}\,
\left(\Pi_{p+M+1} C {A}^{(m)} \slash k\Pi_{M}\right)_{AB}
\nonumber \\
&\times &{\cal
V}^A_{-{1\over 2} + (M+1)}(z)~\tilde{\cal V}^B_{-{3\over 2} -
M}(\bz)~~.\label{brstM1}\\\nonumber
\end{eqnarray}
Therefore, we have
\begin{equation}
\label{din2}
\comm{Q_1\,}{W^{(M)}(z,\bz)} + \comm{\tilde Q_1\,}{W^{(M+1)}(z,\bz)} = 0~~,
\end{equation}
provided the coefficients $a_M$ are given by
\begin{equation}
\label{c13}
a_M = {(-)^{{1\over 2}M(M+1)}\over M! [(M-1)!\ldots 2\cdot 1]^2}~~,
\end{equation}
and the R-R gauge field satisfies
\begin{equation}
\label{c14}
(-)^p \slash k {A}^{(m)} + {A}^{(m)} \slash k = 0~~,
\end{equation}
that is if the potential is in the generalized Lorentz gauge
$d\,*A^{(m)}=0$.
If this is the case, and recalling that
\beq
\comm{{\tilde Q}_1\,}{W^{(0)}(z,\bz)}=0~~,
\label{din3}
\eeq
it follows that the massless vertex (\ref{maa2})
is BRST invariant,
\beq
\comm{Q+\tilde Q\,}{W(z,\bz)} = 0~~.
\label{brstfin}
\eeq
We now show that under a picture changing operation in the right sector
\cite{FMS},
\begin{equation}
\label{casa1}
W(z,\bar z)\to \comm{Q+\tilde Q}{2\tilde \xi(\bar z) W(z,\bar z)}_+~,
\end{equation}
the asymmetric vertex $W$ transforms into the usual operator
$V_\R$ in the symmetric $(-1/2,-1/2)$ picture whose polarization
contains the R-R field strength as shown in \eq{simver}.
To do this, we firstly notice that
since the insertion of $\tilde\xi$ does not alter the conformal
dimension of $W$, the commutation with $\tilde Q_0$ 
is still ensured if $k^2=0$;
secondly, also the commutation with ${\tilde Q_2}$ is not spoiled.
Again, the non-trivial part is the commutator with $\tilde Q_1$.
Utilizing the
OPEs (\ref{c7},\ref{contra3}), we can see that
only the first term $W^{(0)}$ is responsible
for the picture-changing; indeed, we get
\bea
\comm{\tilde Q_1\,}{2\tilde\xi(\bz) W^{(0)}(z,\bar z)}_+ &=&
{1\over 2\sqrt{2}}\left(\Pi_p C{A}^{(m)}\slash k \Pi_1\right)_{AB}
\, {\cal V}^A_{-{1\over 2}}(z) \tilde{\cal V}^B_{-{1\over 2}}(\bar z)
\nonumber \\
&=&V_\R(z,\bar z)~~,
\label{casa2}
\ena
where in the last step we used the Lorentz gauge condition. For the
remaining terms $W^{(M)}$ with $M>0$, the only non-zero contribution
to the picture-changing arises when $\tilde\xi$ acts as a spectator,
so that
\beq
\comm{\tilde Q_1\,}{2\tilde\xi(\bz) W^{(M)}(z,\bz)}_+
=-2\tilde\xi(\bz)\comm{\tilde Q_1\,}{W^{(M)}(z,\bz)}
~~.
\label{pic1}
\eeq
Thus, taking into account Eqs. (\ref{din3}) and
(\ref{brstfin}), we find that
\begin{equation}
\label{casa3}
\comm{Q+\tilde Q\,}{2\tilde\xi(\bz) W(z,\bar z)}_+ = V_\R(z,\bar z) -
2\tilde\xi(\bar z) \comm{Q+\tilde Q\,}{W(z,\bar z)}= V_\R(z,\bz)~~ .
\end{equation}
We conclude this appendix by finding the explicit expression of the state
$\ket{W}$ created by the vertex operator (\ref{maa2}).
For the first term, $W^{(0)}$, we have
by definition
\begin{equation}
 \label{fred1}
 \ket{W^{(0)}} =
 \left(\Pi_p C{\cal A}^{(m)}\Pi_0\right)_{AB} \ket{A;k/2}_{-1/2}
 ~\ket{\tilde B;k/2}_{-3/2}~~.
\end{equation}
Consider now the second term, $W^{(1)}$. We have
\begin{eqnarray}
 \label{fred2}
 \ket{W^{(1)}} & = &
 -\left(\Pi_{p+1} C{A}^{(m)}\Pi_1\right)_{AB}
 \ket{A;k/2}~\ket{\tilde B;k/2}
 \nonumber\\
 & \times & \lim_{z\to 0} \eta(z) \ket{1/2}_\phi~
 \lim_{\bar z\to 0} \bar\partial\tilde\xi(\bar z) 
\ket{-5/2}_{\tilde\phi}~~,
 \end{eqnarray}
where the notation $\ket{\ell}_\phi$ stands for the vacuum of the $(\phi,\eta,\xi)$
system with $\phi$-momentum $\ell$, which, according to \eq{picture}, coincides
with the superghost vacuum in the $P=\ell$ picture.
Keeping this in mind, and using the fermionization formulas
(\ref{bosoniz}), it is easy to see that
\begin{eqnarray}
 \label{fred3}
 &&\lim_{z\to 0} \eta(z)
 \ket{1/2}_\phi ~=~
 -\lim_{z\to 0} z^{-{1\over 2}}\,\gamma(z) \ket{P=-1/2}
~=~ -\gamma_0\ket{P=-1/2}~~,
\nonumber\\
 &&\lim_{z\to 0}\bar\partial\tilde\xi(\bar z)
 \ket{-5/2}_{\tilde\phi} ~=~
  \lim_{\bar z\to 0} \bar z^{{3\over 2}}\,\tilde\beta(\bar z)
\ket{\tilde P=-3/2} ~=~
 \tilde\beta_0\ket{\tilde P=-3/2} ~~,
\end{eqnarray}
so that
\begin{equation}
 \label{fred4}
 \ket{W^{(1)}} =
  \left(\Pi_{p+1} C{\cal A}^m\Pi_1\right)_{AB}\, \gamma_0 \tilde\beta_0\,
 \ket{A;k/2}_{-1/2}\,
 \ket{\tilde B;k/2}_{-3/2}~~.
\end{equation}
It is clear that all terms of $W$ with $M$ even (odd) will give exactly the same
structure of $\ket{W^{(0)}}$ ($\ket{W^{(1)}}$), except for the superghost part.
To deal with it, we use the generalization of
the formulas in \eq{fred3}, namely
\begin{eqnarray}
 \label{fred5}
 \lim_{z \to 0} \partial^{M-1} \eta (z)\ldots
 \eta (z)
 \ket{-1/2+M}_\phi & = &
 (-)^{M(M+1)\over 2}\, (M-1)!\ldots 2!\, \gamma_0^M\,
 \ket{P=-1/2}~,\nonumber\\
 \lim_{\bar z \to 0} \bar\partial^M \tilde\xi (\bar z)
 \ldots\bar\partial\tilde\xi(\bar z)
 \ket{-3/2 - M}_{\tilde\phi} & = &
 (-)^{M(M-1)\over 2}\, (M-1)!\ldots 2!\, \tilde\beta_0^M
 \ket{\tilde P=-3/2}~.\nonumber\\
\end{eqnarray}
Then, using the explicit expression of $W^{(M)}$, \eq{c10} and
\eq{c13}, we finally obtain
\begin{eqnarray}
\label{fred6}
\ket{W} & = & \left\{
\left(\Pi_p C{\cal A}^{(m)}\Pi_0\right)_{AB}\,
\cos (\gamma_0\tilde\beta_0) +
\left(\Pi_{p+1} C{\cal A}^{(m)}\Pi_1\right)_{AB}\, \sin 
(\gamma_0\tilde\beta_0) \right\}\nonumber\\
& \times & \ket{A;k/2}_{-1/2}~\ket{\tilde B;k/2}_{-3/2}~~.
\end{eqnarray}
Converting to the Majorana-Weyl notation, one can check that
the state (\ref{fred6}) for a Type IIA theory coincides with \eq{-3/2state}.
%%%%%%%%%%%%%%%%%%%%%%%%%%%
\vskip1cm
\appendix{\Large {\bf {Appendix B}}}
\label{appb}
\vskip 0.5cm
\renewcommand{\theequation}{B.\arabic{equation}}
\setcounter{equation}{0}
\noindent
In this appendix we prove the BRST equivalence between the on-shell 
R-R vertex operators $W_G$ and $W$ defined respectively
in Eqs. (\ref{WG}) and (\ref{Rvertex}). For definiteness,
we study in detail vertices associated to a 1-form
potential, but of course our results are completely general.
Let us then consider the vertex $W_G$ for a 1-form, which
explicitly reads
\begin{equation}
{W}_G (k;z,\bz) =\frac{1}{2}
\left[ A_\mu(C\Gamma^\mu)_{\a\b} 
+ \frac{1}{3!} A_{\mu \nu \rho}
(C\Gamma^{\mu \nu \rho})_{\alpha \beta}\right]
{\cal V}_{-1/2}^{\a} (z)
{\tilde{\cal V}}_{-3/2}^{\beta}(\bz)~~,
\label{WS}
\end{equation}
where ${\cal V}_{\ell}^A$ is defined in \eq{genefo3}. This vertex 
is BRST invariant if (see \eq{scond})
\begin{eqnarray}
k^\mu A_\mu  =  0~~,~~   
k_\mu A_\nu - k_\nu A_\mu + k^\rho A_{\rho \mu \nu} = 0
~~,~~k_{[\lambda} A_{\mu \nu \rho ]} = 0 ~~.
\label{potc}
\end{eqnarray}
These conditions imply that $k^2=0$, and that the 3-form
potential is pure gauge.

We now show that by adding to \eq{WS} a BRST exact operator
we can reconstruct the vertex $W$ with indefinite left and right
superghost numbers. To this aim, let us
begin by considering the following operator
\begin{equation}
{\Lambda}^{(0)}(k;z,\bz) = \sqrt{2}
\Lambda_{\mu\nu}\left(C\Gamma^{\mu\nu}\right)_{\a \dot{\beta}}
{\cal{V}}_{-1/2}^{\alpha} (z) {\tilde{\cal{V}}}_{-5/2}^{\dot{\beta}}
({\bar{z}}) ~~,
\label{w0}
\end{equation}
where $k^2=0$ and 
$\Lambda_{\mu\nu}$ is a 2-form to be specified later. 
It is simple to show that $\Lambda^{(0)}$ commutes with
$Q_0$, $\tilde Q_0$, $Q_2$ and $\tilde Q_2$,
whereas the commutators with $Q_1$ and $\tilde Q_1$ are not
vanishing and read
\bea
\comm{Q_1\,}{\Lambda^{(0)}(k;z,\bz)}
&=&
\frac{1}{4}\Lambda_{\mu\nu}\left(C\slash{k}\Gamma^{\mu\nu}
\right)_{{\dot{\alpha}}{\dot{\beta}}}
{\cal{V}}_{+1/2}^{{\dot\alpha}} (z)
{\tilde{\cal{V}}}_{-5/2}^{\dot{\beta}}
(\bz)~~,
\label{qw0}
\\
\comm{\tilde Q_1\,}{\Lambda^{(0)}(k;z,\bz)}
&=&
\frac{1}{4}\Lambda_{\mu\nu}\left(C\Gamma^{\mu\nu}\slash{k} 
\right)_{\a \b }
{\cal{V}}_{-1/2}^{\a } (z)
{\tilde{\cal{V}}}_{-3/2}^{\b }
(\bz)~~.
\label{qw1}
\ena
The right hand side of \eq{qw1} has the same operator structure of
$W_G$ and also of the first term of $W$, {\em i.e.} $W^{(0)}$, whereas 
the right hand side of \eq{qw0} has the same operator structure of 
$W^{(1)}$. Thus, we can reach our goal by simply adjusting the 
polarization coefficients. In fact, if we choose $\Lambda_{\mu\nu}$
such that
\begin{equation}
\Lambda_{\mu\nu}k^\nu = A_\nu~~,
\label{lam}
\end{equation}
it follows that
\beq
\comm{\tilde Q_1\,}{\Lambda^{(0)}(k;z,\bz)}
=\frac{1}{2}
\left[ A_\mu \left(C \Gamma^\mu \right)_{\alpha \beta}
- \frac{1}{3!} A'_{\mu \nu \rho}
\left(C \Gamma^{\mu \nu \rho} \right)_{\alpha \beta}
\right]
{\cal V}_{-1/2}^{\a} (z)\,
{\tilde{\cal V}}_{-3/2}^{\beta}(\bz)~,
\label{WS1}
\end{equation}
where $A'_{\mu\nu\rho}=-k_{[\mu}\Lambda_{\nu\rho]}$.
Since this 3-form satisfies exactly the same properties as 
$A_{\mu\nu\rho}$ in \eq{potc}, we can identify them
so that we have 
\bea
W_G(k;z,\bz) + \comm{\tilde Q_1\,}{\Lambda^{(0)}(k;z,\bz)}
&=& A_\mu \left(C \Gamma^\mu \right)_{\alpha \beta}
{\cal V}_{-1/2}^{\a} (z)\,
{\tilde{\cal V}}_{-3/2}^{\beta}(\bz)
\nonumber \\
&=& W^{(0)}(k;z,\bz)~~.
\label{1term}
\ena
Let us now consider the following operator
\begin{equation}
{\Lambda}^{(1)}(k;z,\bz) = -\frac{\sqrt{2}}{2}
\Lambda_{\mu\nu}\left(C\Gamma^{\mu\nu}\right)_{\dot\a {\beta}}
{\cal{V}}_{1/2}^{\dot \alpha} (z) \,{\tilde{\cal{V}}}_{-7/2}^{\beta}
({\bar{z}}) ~~,
\label{l1}
\end{equation}
with $k^2=0$, which  commutes with
$Q_0$, $\tilde Q_0$, $Q_2$ and $\tilde Q_2$. Then, using 
Eqs. (\ref{lam}) and (\ref{qw0}), straightforwardly enough we get
\bea
\comm{Q_1\,}{\Lambda^{(0)}(k;z,\bz)}
+\comm{\tilde Q_1\,}{\Lambda^{(1)}(k;z,\bz)}
&=& -A_{\mu}\left(C\Gamma^{\mu}
\right)_{{\dot{\alpha}}{\dot{\beta}}}
{\cal{V}}_{+1/2}^{{\dot\alpha}} (z)
{\tilde{\cal{V}}}_{-5/2}^{\dot{\beta}}
(\bz)
\nonumber \\
&=& W^{(1)}(k;z,\bz) ~~.
\label{2term}
\ena
This procedure can be iterated because it is always possible
to find operators ${\Lambda}^{(M)}(k;z,\bz)$ such that
\beq
\comm{Q_1\,}{\Lambda^{(M-1)}(k;z,\bz)}
+\comm{\tilde Q_1\,}{\Lambda^{(M)}(k;z,\bz)}
= W^{(M)}(k;z,\bz)
\label{Mterm}
\eeq
for any $M\geq 0$. Combining all these equations, we finally obtain
\beq
W_G(k;z,\bz) + \comm{Q_1+\tilde Q_1\,}{\Lambda(k;z,\bz)}
= W(k;z,\bz)~,
\label{fin}
\eeq
where $\Lambda(k;z,\bz)=\sum_M\Lambda^{(M)}(k;z,\bz)$.
This equation implies that the states created by $W_G$ and $W$ are
related to each other as shown in \eq{equivRS}. 
The states $\ket{W}$ and $\ket{W_G}$ are BRST equivalent only on 
shell, while if $k^2\not =0$ they are not related even in the restricted 
cohomology.
%%%%%%%%%%%%%%%%%%%
\vskip 1cm
%\vfill

\end{document}